\documentclass[11pt, preprint]{aastex}
\usepackage{subfigure}
\let\tablehead\undefined
\let\tabletail\undefined
\usepackage{supertabular}
\usepackage{rotating}
\let\affil\undefined
\usepackage{authblk}
\title{\bf The McDonald Observatory Planet Search: New Long-Period Giant Planets, and Two Interacting Jupiters in the HD 155358 System}
\author[1]{Paul Robertson}
\author[1]{Michael Endl}
\author[1]{William D. Cochran}
\author[1]{Phillip J. MacQueen}
\author[2]{Robert A. Wittenmyer}
\author[2]{J. Horner}
\author[1]{Erik J. Brugamyer}
\author[3,4]{Attila E. Simon}
\author[1]{Stuart I. Barnes}
\author[1]{Caroline Caldwell}
\affil[1]{Department of Astronomy and McDonald Observatory, University of Texas at Austin, Austin, TX 78712, USA; paul@astro.as.utexas.edu}
\affil[2]{Department of Astrophysics and Optics, School of Physics, University of New South Wales, Sydney, NSW 2052, Australia}
\affil[3]{Konkoly Observatory of the Hungarian Academy of Sciences, PO. Box 67, H-1525 Budapest, Hungary}
\affil[4]{Department of Experimental Physics and Astronomical Observatory, University of Szeged, 6720 Szeged, Hungary}

\begin{abstract}
We present high-precision radial velocity (RV) observations of four solar-type (F7-G5) stars -- HD 79498, HD 155358, HD 197037, and HD 220773 -- taken as part of the McDonald Observatory Planet Search Program.  For each of these stars, we see evidence of Keplerian motion caused by the presence of one or more gas giant planets in long-period orbits.  We derive orbital parameters for each system, and note the properties (composition, activity, etc.) of the host stars.  While we have previously announced the two-gas-giant HD 155358 system, we now report a shorter period for planet c.  This new period is consistent with the planets being trapped in mutual 2:1 mean-motion resonance.  We therefore perform an in-depth stability analysis, placing additional constraints on the orbital parameters of the planets.  These results demonstrate the excellent long-term RV stability of the spectrometers on both the Harlan J. Smith 2.7 m telescope and the Hobby-Eberly telescope.
\end{abstract}


\begin{document}

\section{\bf Introduction}

Beginning with the first radial velocity exoplanet detections \citep{latham89,mayor95}, exoplanet surveys have identified a large number of gas giant planets on very short orbits \citep[see][for complete details]{udry07,borucki11,wright11}.  However, to answer the fundamental question ``how common are planetary systems analogous to our solar system?'' requires years of data from dedicated exoplanet surveys to achieve the observational time baseline necessary to detect the true Jupiter analogs--those giant planets which do not undergo significant inward migration after the dissipation of the protoplanetary disk.  Planets with periods greater than approximately 2 years are particularly valuable, as they are outside the detection limits of space-based transit searches such as \emph{Kepler} \citep{borucki10} and CoRoT \citep{baglin03}, whose mission timescales are too short to confirm such detections.  

Identifying long-period Jovian planets is essential to constraining theoretical and observational results on the true distribution of giant planets versus orbital separation.  \citet{cumming08} find approxmiately 10.5\% of FGK stars host gas giant planets with periods between 2 and 2000 days, and \citet{witt11} claim a 3.3\% occurrence rate for "Jupiter analogs," which they define as giant planets with $e < 0.2$ and $P \ge 8$ years.  While microlensing surveys \citep[e.g.][]{gould10} derive a higher giant planet fraction, their results are roughly consistent with RV surveys if we consider the differences in sensitivity between the two methods.  Microlensing planet searches are particularly sensitive to planets near the Einstein radius (typically 2-4 AU), where core accretion models \citep{benz08} and extrapolations of RV results \citep{cumming08} agree there should exist a significant population of Jovian planets.  The expansion of the time baselines of RV surveys to include these long-period planets is essential to reconcile the statistics of RV and microlensing planet searches.    

A more complete census of long-period gas giants also places strong constraints on theories of planet formation and migration.  The frequency of Jovian planets offers information on the efficiency of core accretion in protoplanetary disks \citep{sdr09,mann10,bromley11}, and the mass-period distribution will test predictions that planets with $m \sin i \ge M_J$ do not migrate as far as Neptune- or Saturn-mass planets \citep{delp05,bromley11}.  Additionally, enlarging the sample size of giant planets allows for a more robust examination of correlations between the properties of exoplanetary systems (e.g. planet frequency, mass, eccentricity) and those of their stellar hosts, such as metallicity \citep{fischer05}, mass \citep{johnson11}, and galactic dynamics \citep{ecuvillon07}.

The McDonald Observatory Planetary Search \citep{cochran93} has been conducting a high-precision radial velocity survey to identify substellar companions around FGK stars with the Harlan J. Smith 2.7 m telescope since 1987.  Since our migration to a cross-dispersed echelle spectrograph in combination with an I$_2$ absorption cell in 1998, and with the addition of the Hobby-Eberly Telescope in 2001, we have successfully monitored hundreds of stars with velocity precision of $\sim 3$ m/s, giving us a 14 year observational time baseline.  As a result, we now see evidence of planets with periods of 5 years or more, a demographic that is still underrepresented in exoplanet discoveries, accounting only 38 of the 518 planets listed in the exoplanets.org database (as of 5 January 2012).  In this paper, we present two such objects--HD 79498b and HD 220773b--a  shorter-period Jovian planet around HD 197037, and updated orbital parameters for the two-planet system surrounding HD 155358.

\section{\bf Sample and Observations}

The McDonald Observatory Planet Search Program currently monitors over 250 (mostly FGK) stars with the 2.7 m Harlan J. Smith Telescope for RV variations due to planetary companions.  The survey is magnitude limited to $V \sim 10$, and regularly achieves RV precision of 3-6 m/s.  The large number of nights provided for our program results in excellent temporal coverage and long observational time baselines for our targets; the objects presented here each have 30-120 data points over 7-10 years.  Thus, our sensitivity is more than sufficient to detect RV signatures comparable to that of Jupiter, which would require observations at our level of precision over 12 years.

\subsection{2.7 m Telescope Observations}

The velocities for HD 79498 and HD 197037 were obtained with the Smith Telescope's Tull Coud\'{e} Spectrograph \citep{tull95} using a $1.8 \arcsec$ slit, giving a resolving power of R = 60,000.  Before starlight enters the spectrograph, it passes through an absorption cell containing iodine (I$_{2}$) vapor maintained at 50$^{\circ}$ C, resulting in a dense forest of molecular absorption lines over our stellar spectra from 5000-6400 \AA.  These absorption lines serve as a wavelength metric, allowing us to simultaneously model the instrumental profile (IP) and radial velocity at the time of the observation.  For each star, we have at least one high-S/N iodine-free template spectrum, which we have deconvolved from the IP using the Maximum Entropy Method, and against which the shifts due to the star's velocity and the time-variant IP are modeled.  Our reported RVs are measured relative to this stellar template, and are further corrected to remove the velocity of the observatory around the solar system barycenter.  All radial velocities have been extracted with our pipeline AUSTRAL \citep{endl00}, which handles the modeling of both the IP and stellar velocity shift.

\subsection{HET Observations}

Our RV data for HD 155358 and HD 220773 were taken with the High Resolution Spectrometer \citep[HRS;][]{tull98} on the queue-scheduled 9.2m Hobby-Eberly Telescope \citep[HET;][]{ramsey98}.  As with the 2.7 m observations, the HET/HRS spectra are taken at R = 60,000 with an I$_{2}$ absorption cell.  The fiber-fed HRS is located below the telescope in a temperature-controlled room.  Separate stellar templates were obtained with HRS for these stars.  Details of our HET observing procedure are given in \citet{cochran04}.

While we use HRS spectra exclusively for obtaining RVs for HD 155358 and HD 220773, we have I$_2$-free spectra from the 2.7 m telescope for these stars as well.  These spectra serve two purposes.  First, they allow us to determine the stellar parameters for all five stars using the same instrumental setup.  Also, the Tull 2.7 m coud\'{e} spectrograph provides Ca H and K indices, which contain information as to the activity levels of the stars.

Tables \ref{79498tab}-\ref{220773tab} list the relative velocities and their associated uncertainties for HD 79498, HD 155358, HD 197037 and HD 220773, respectively.  Table \ref{155358tab} includes observations published in \citet{cochran07}, but since all of our spectra have been re-reduced with our most up-to-date methods, the velocities presented here have a higher precision.

\section{\bf Analysis and Orbit Modeling}

\subsection{Host Star Characterization}

We determine the stellar parameters of our targets according to the procedure described in \citet{brugamyer11}.  The method relies on a grid of ATLAS9 model atmospheres \citep{kurucz93} in combination with the local thermodynamic (LTE) line analysis and spectral synthesis program MOOG\footnote{available at http://www.as.utexas.edu/$\sim$chris/moog.html} \citep{sneden73}.  Using the measured equivalent widths of 53 neutral iron lines and 13 singly-ionized iron lines, MOOG force-fits elemental abundances to match the measured equivalent widths according to built-in atomic line behavior.  Stellar effective temperature is determined by removing any trends in equivalent widths versus excitation potential, assuming excitation equilibrium.  Similarly, we compute the stellar microturbulent velocity  $\xi$ by eliminating trends with reduced equivalent width ($\equiv$ W$_{\lambda}$/$\lambda$).  Finally, by assuming ionization equilibrium, we constrain stellar surface gravity by forcing the abundances derived from Fe I and Fe II lines to match.

We begin our stellar analysis by measuring a solar spectrum taken during daylight (through a solar port) with the same instrumental configuration used to observe our targets.  The above procedure yields values of T$_{eff} = 5755 \pm 70$ K, $\log g = 4.48 \pm 0.09$ dex, $\xi = 1.07 \pm 0.06$ km/s, and $\log \epsilon ($Fe$) = 7.53 \pm 0.05$ dex.  We then repeat this analysis for our target stars, using the I$_2$-free template spectra.  We note that our derived metallicities are, as is conventional, differential to solar.  The high S/N and spectral resolution of our stellar templates allow us to make robust estimates of each star's effective temperature, $\log g$, metallicity, and microturbulent velocity, which we include in Table \ref{stellar}.  We also include photometry, parallax data, and spectral types from the ASCC-2.5 catalog \citep[Version 3,][]{kharchenko09}, as well as age and mass estimates from \citet{casagrande11}.

In addition to basic stellar parameters, we use the standard Ca H and K indices to evaluate the hypothesis that our observed RV signals are actually due to stellar activity.  For HD 79498 and HD 197037 we have time-series measurements of the Mount Wilson $S_{HK}$ index.  For each of these stars, we obtain $S_{HK}$ simultaneously with radial velocity, so we include those values in the RV tables.  From the average value of these measurements, we derive \citep[via][]{noyes84} $\log R'_{HK}$, the ratio of Ca H and K emission to the integrated luminosity of the star, which we include in Table \ref{stellar}.  The activity indicators are discussed in detail for each star below, but we note here that the sample overall appears to be very quiet, and we have no reason to suspect stellar variability as the cause of the observed signals. 

As a second fail-safe against photospheric activity mimicking Keplerian motion, we calculate the bisector velocity spans (BVS) for spectral lines outside the I$_2$ absorption region.  As described in \citet{brown08}, this calculation offers information regarding the shapes of the stellar absorption lines in our spectra.  As photospheric activity that might influence RV measurements occurs, it alters the shapes of these lines, causing a corresponding shift in the BVS.  Our BVS, then, offer a record of the activity of our target stars.  Provided our planetary signals are real, we expect the BVS to be uncorrelated with our RV measurements.  For each RV point, we have a corresponding BVS measurement computed from the average of the stellar lines outside the I$_2$ region.

\subsection{Orbit Fitting}

To determine the orbital parameters of each planetary system, we first analyze each RV set using the fully generalized Lomb-Scargle periodogram \citep{zechmeister09}.  We estimate the significance of each peak in the periodogram by assigning it a preliminary false-alarm probability (FAP) according to the method described in \citet{sturrock10}.  To conclude our periodogram analysis, we assign a final FAP for each planet using a bootstrap resampling technique.  Our bootstrap method, which is analogous to the technique outlined in \citet{kurster97} retains the original time stamps from the RV data set, and selects a velocity from the existing set (with replacement) for each observation.  We then run the generalized Lomb-Scargle periodogram for the resampled data.  After 10,000 such trials for each data set,  the FAP is taken to be the percentage of resampled periodograms that produced higher power than the original RV set.  Power in the \citet{zechmeister09} periodogram is given by $\Delta\chi^2 \equiv \chi_{0}^{2} - \chi_{P}^{2}$, or the improvement of fit of an orbit at period $P$ over a linear fit, so a ``false positive'' result in the bootstrap trial occurs when a random sampling of our measured RVs is better suited to a Keplerian orbit than the actual time series.  In Table \ref{orbits}, we include both FAP estimates for comparison, but we adopt this bootstrap calculation as our formal confidence estimate.

The periods identified in the periodograms are then used as initial estimates for Keplerian orbital fits, which we perform with the GaussFit program \citep{jefferys88}.  We list the orbital parameters of each planet in Table \ref{orbits}, and describe the individual systems in detail below.  As a consistency check, we also fit each orbit using the SYSTEMIC console \citep{meschiari09}, finding good agreement between our results in every case.  An additional advantage of the SYSTEMIC console is that it computes a stellar ``jitter'' term--a measure of random fluctuations in the stellar photosphere--\citep[e.g.][]{queloz01} for each star, which we include in Table \ref{orbits}.

\section{\bf New Planetary Systems}

\subsection{HD 79498}

\subsubsection{RV Data and Orbit Modeling}

\begin{figure}
  \begin{center}
    \includegraphics[scale=0.6]{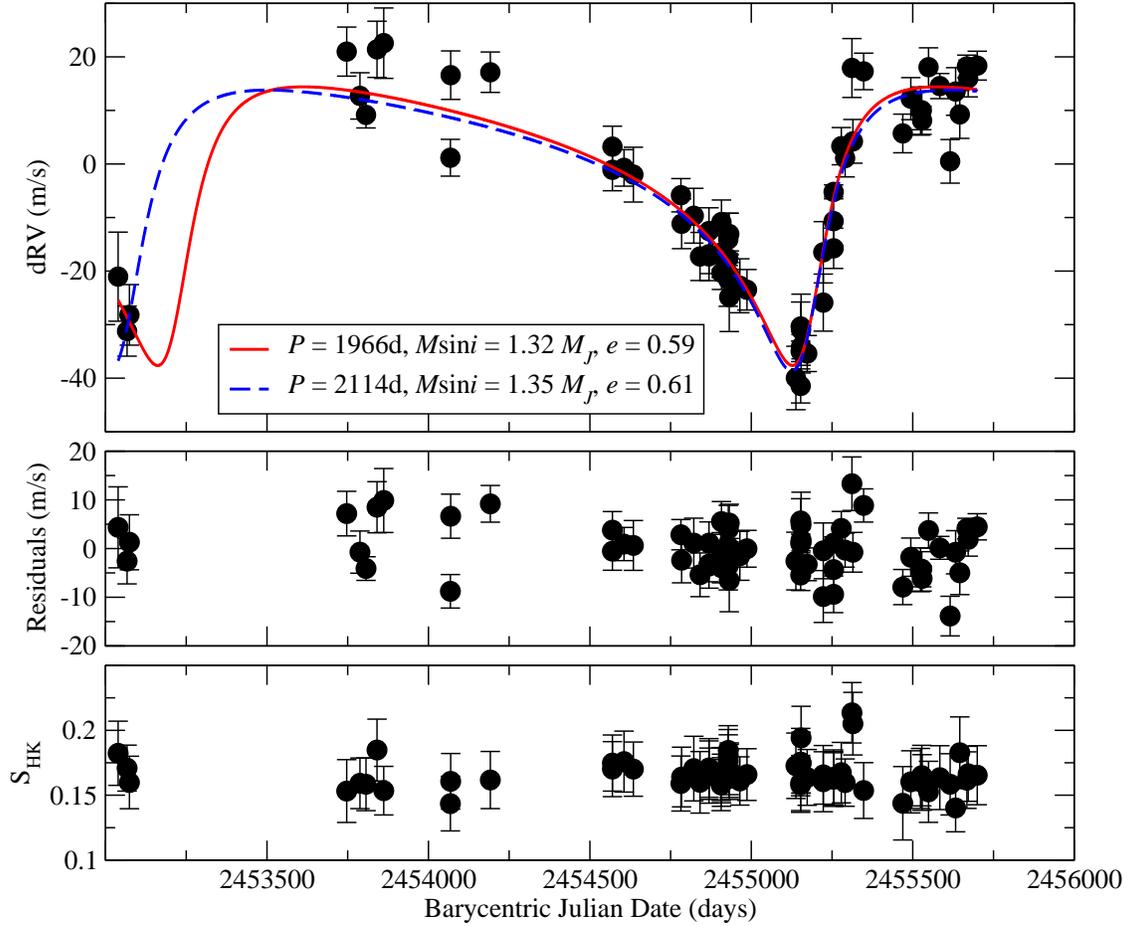}
    \caption{\emph{Top}: Radial velocity data for HD 79498.  The best-fit orbit model is shown as a solid red line.  Our second acceptable fit is also included, shown as a dashed blue line.  \emph{Middle}: Residual RVs after subtraction of a one-planet fit.  \emph{Bottom}: $S_{HK}$ as measured at each RV point.}
    \label{79498rv}
    \end{center}
\end{figure}

Our RV data for HD 79498 consist of 65 observations taken over 7 years from the 2.7 m telescope.  The data have an RMS scatter of 18.3 m/s and a mean error of 4.15 m/s, indicating significant Doppler motion.  Our RVs are listed in Table \ref{79498tab}, and plotted in Figure \ref{79498rv}.

The periodogram for the HD 79498 RVs reveals a large peak around 1815 days, with a preliminary FAP of $4.09 \times 10^{-9}$.  Performing a Keplerian fit with $P = 1815$ days as an initial guess, we find a single-planet solution with the parameters $P = 1966$ days, $e = 0.59$, and $K = 26.0$ m/s, indicating a planetary companion with $M \sin i = 1.34 M_{J}$ at $a = 3.13$ AU.  Considering the high eccentricity and the width of the periodogram peak, these values are in good agreement with our initial guess.  This fit produces a reduced  $\chi^{2} = 1.77$ with an RMS scatter of 5.13 m/s around the fit.  We note that we have not included the stellar ``jitter'' term in our error analyses, but we have verified that most of the $\chi^2$ excess above unity can be attributed to our fitted value of 2.76 m/s.  In a series of 10$^4$ trials of our bootstrap FAP analysis, we did not find an improvement in $\Delta\chi^2$ for any resampled data set, resulting in an upper limit of $10^{-4}$ for the FAP of planet b.  Figure \ref{79498rv} shows our fit, plotted over the time-series RV data.  We include the full parameter set for HD 79498b in Table \ref{orbits}.

Our analysis of HD 79498b admits a second, slightly different orbital solution of nearly equal significance.  This solution converges at $P = 2114$ days, $e = 0.61$, and $K = 26.2$ m/s.  The corresponding planetary parameters then become $M \sin i = 1.35 M_{J}$ and $a = 3.27$ AU.  For comparison, we include the plot of this fit in Figure \ref{79498rv}.  Although the shorter-period solution may appear to be driven mainly by the first data point, we note that our fitting routines converge to the 1966-day period regardless of whether that point is included, and that the fitting statistics are still better for the shorter period with the point excluded.  Clearly, the qualitative properties of the planet remain unchanged regardless of the choice of parameters.  However, because of the slightly better fitting statistics ($\chi^{2} = 1.77$ versus $\chi^{2} = 1.84$, RMS = 5.13 m/s versus RMS = 5.50 m/s) for the first model, we adopt it as our formal solution.

We computed the periodogram of the residuals around the fit of HD 79498b to search for additional signals.  We see no evidence of additional planets in the system.  We note that no additional signals appear in the residuals of the alternate fit for planet b either.

\subsubsection{Stellar Activity and Line Bisector Analysis}

With a $\log R'_{HK} = -4.66$, HD 79498 appears to be a low-activity star, and our indicators corroborate that notion.  Our line bisector velocity spans are well behaved, with an RMS scatter of 19 m/s, below the $K$ amplitude of planet b.  The BVS are uncorrelated with the measured RVs, and a periodogram reveals no periodicity in the bisector velocities.  Likewise, the $S_{HK}$ time series shows no significant signals, and no correlation with the RV series.  All measures suggest that HD 79498 is a very quiet star, and should neither mask nor artificially produce large-amplitude RV signals such as the one discussed above.

\subsubsection{Stellar Companions}

HD 79498's location on the sky places it near two faint ($V = 10, 11$) stars, each separated by approximately 60 arcseconds \citep{dommanget02}.  When examining the proper motions of all three stars, though, we see that only the southern companion, BD+23 2063B, is actually associated with HD 79498, making it a double star system \citep{bonnarel00}.  At a distance of 49 parsecs, the companion star is located at a minimum distance $d \sim 2900$ AU from HD 79498.  Given an overly generous mass estimate of $M = 1 M_{\odot}$, the secondary star would impart a gravitational acceleration of just $GM_{\odot}/d^2 = 0.022$ m s$^{-1}$ yr$^{-1}$, well below the sensitivity of our instrument.  Furthermore, at 2900 AU, the orbital period of the companion would be sufficiently long to appear as a linear slope in our data, not the orbit discussed above.  It is possible, however, that the presence of this distant star may excite Kozai cycles \citep{kozai62,lidov62} in HD 79498b, which serve to maintain its high eccentricity \citep[see, e.g.][]{katz11,witt07}.

\subsection{HD 197037}

\subsubsection{RV Data and Orbit Modeling}

\begin{figure}
  \begin{center}
    \includegraphics[scale=0.6]{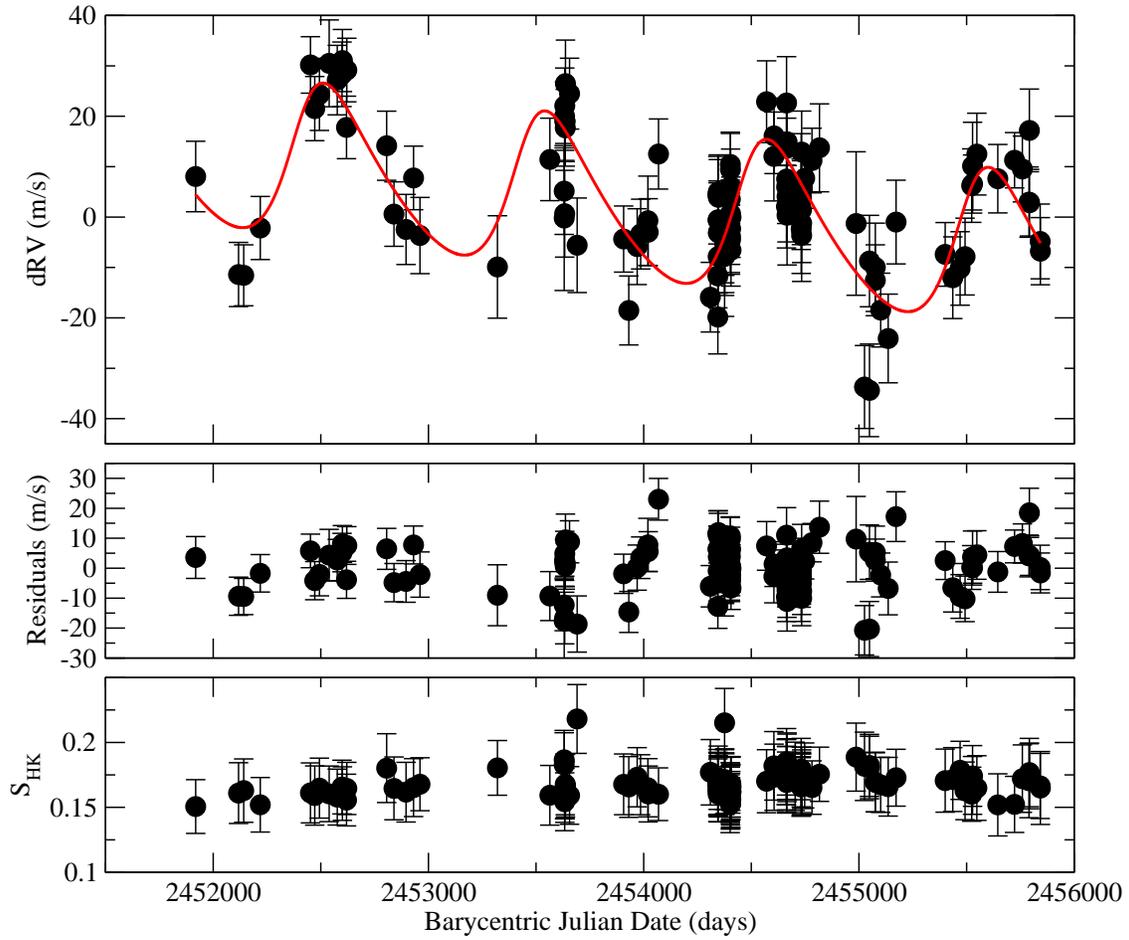}
    \caption{\emph{Top}: Radial velocity data for HD 197037.  The best-fit orbit model is shown as a solid red line.  \emph{Middle}: Residual RVs after subtraction of a one-planet fit and a linear trend.  \emph{Bottom}: $S_{HK}$ as measured at each RV point.}
    \label{197037rv}
    \end{center}
\end{figure}

We have obtained 113 RV points for HD 197037 over 10 years from the 2.7 m telescope.  These data have an RMS scatter of 13.1 m/s and a mean error of 7.65 m/s.  We report our measured RVs in Table \ref{197037tab}, and plot them as a time series in Figure \ref{197037rv}.

\begin{figure}
  \begin{center}
    \subfigure[\label{197037ps}]{\includegraphics[scale=0.45]{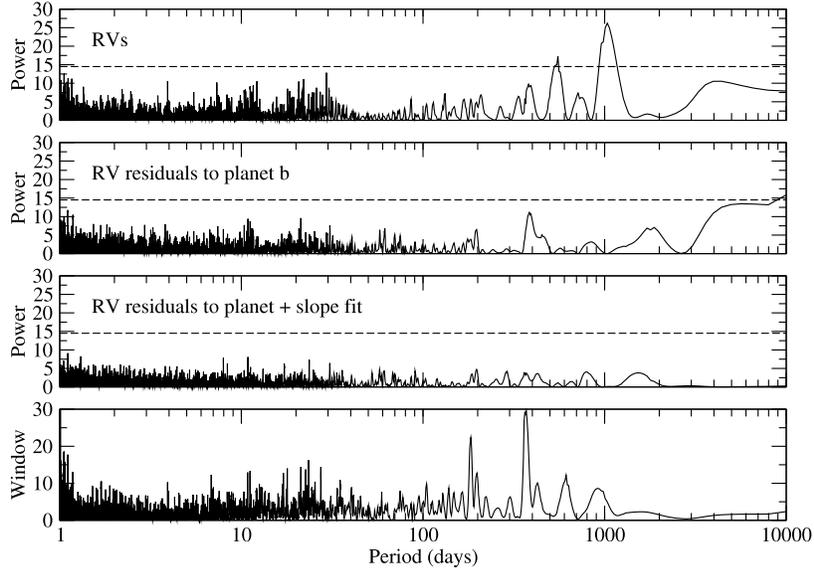}}
    \subfigure[\label{197037bvsps}]{\includegraphics[scale=0.45]{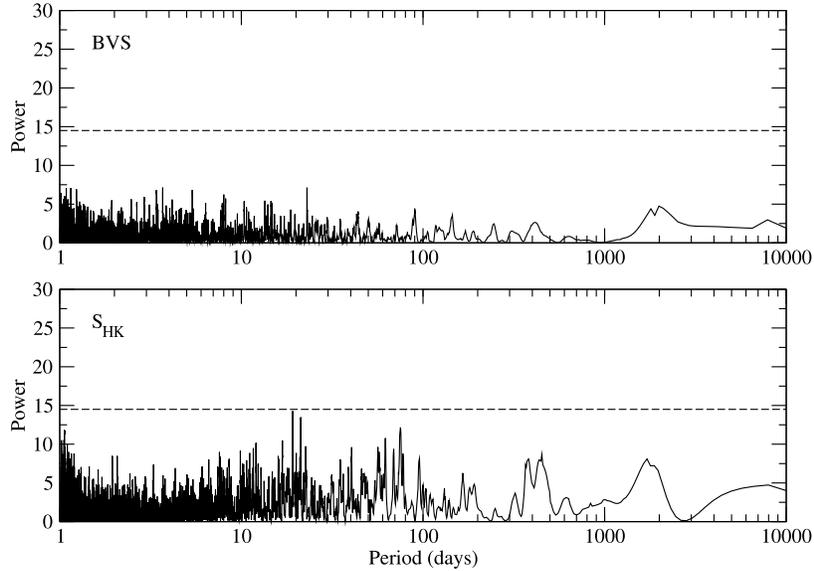}}
    \caption{a.  \emph{From top}: [\emph{1}] Generalized Lomb-Scargle periodogram for HD 197037 RVs.  [\emph{2}] The same periodogram for the residual RVs after subtracting a one-planet fit.  [\emph{3}] Periodogram of the residual RVs after subtracting a one-planet fit and a linear trend.  [\emph{4}] Periodogram of our time sampling (the window function).  
\newline
\newline
b.  Generalized Lomb-Scargle periodograms for the BVS (\emph{top}) and $S_{HK}$ indices (\emph{bottom}) of our spectra for HD 197037.  The dashed lines indicate the approximate power level for a FAP of 0.01.}
    \end{center}
\end{figure}

Our periodogram analysis of HD 197037, shown in Figure \ref{197037ps}, indicates a strong peak around $P \sim 1030$ days.  The power in this peak corresponds to a preliminary FAP of $8.53 \times 10^{-8}$.  Our one-planet Keplerian model yields parameters of $P = 1036$ days, $e = 0.22$, and $K = 15.5$ m/s, showing excellent agreement with the prediction of our periodogram.  The inferred planet has a minimum mass of $0.79 M_{J}$ and lies at an orbital separation of 2.07 AU.  We overplot our model with the RVs in Figure \ref{197037rv}.

The addition of a single Keplerian orbit only reduces the residual RMS of our RVs to 9.18 m/s, which is nearly a factor of two worse than our other one-planet fits.  Furthermore, the periodogram of the residual RVs (Figure \ref{197037ps}) shows a significant increase in power at long periods.  While any additional periods are too far outside our observational time baseline to properly evaluate, we can generate a preliminary two-planet fit with a second $\sim 0.7 M_{J}$ planet with a period around 4400 days and an eccentricity 0.42.  However, our current RV set can also be modeled as a single planet and a linear trend with slope $-1.87 \pm 0.3$ m s$^{-1}$ yr$^{-1}$.  Both fits give a $\chi^2$ of 1.10 and an RMS scatter of 8.00 m/s, so we adopt the more conservative planet-plus-slope model pending further observations.  HD 197037 has no known common proper motion companions within 30 arcseconds, so it is very possible that we are seeing evidence for a distant giant planet or brown dwarf companion.  The final orbital parameters reported in Table \ref{orbits} are derived from the planet+slope model.  We note that we see no additional signals in the residuals of either the two-planet or planet-slope fits.

\subsubsection{Stellar Activity and Line Bisector Analysis}

\begin{figure}
  \begin{center}
    \includegraphics[scale=0.6]{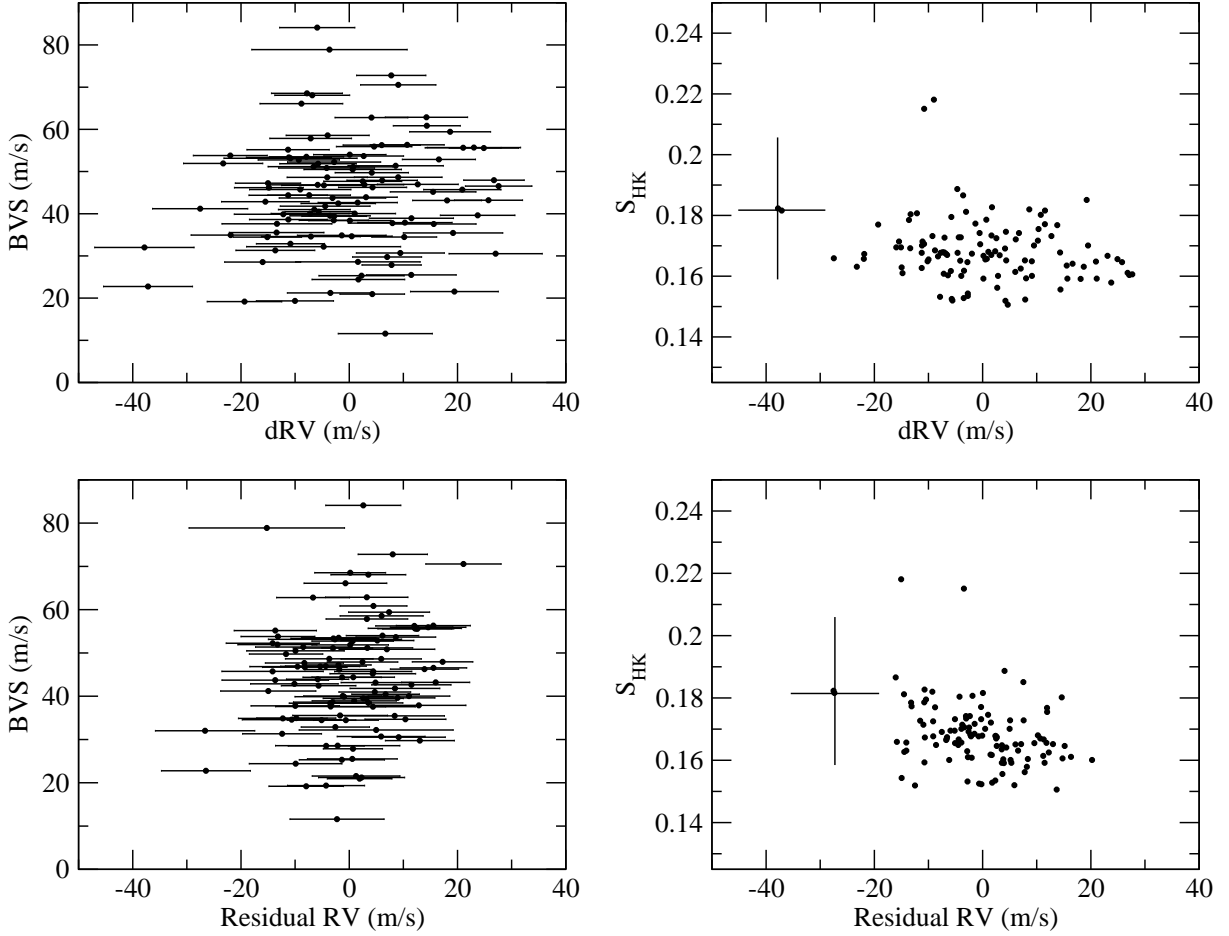}
    \caption{\emph{Left}: Bisector velocity spans plotted against our measured RVs (\emph{top}) and residual RVs to a one-planet fit (\emph{bottom}) for HD 197037.  \emph{Right}: $S_{HK}$ indices plotted against our measured RVs (\emph{top}) and residual RVs to a one-planet fit (\emph{bottom}) for HD 197037.  The error bars shown are representative of the data set.}
    \label{197037bvs}
  \end{center}
\end{figure}

Even after accounting for the linear trend in the residuals around our fit to planet b, the scatter in our RVs is still higher than we typically expect from our 2.7 m data.  Because HD 197037 is an earlier type (F7) than the other stars discussed in this paper, we expect a lower precision as a result of fewer spectral lines to determine velocities. Additionally, with $\log R'_{HK} = -4.53$, it is the most active of the stars presented here.  With an RMS scatter in the bisector velocity spans of 13 m/s, we must take particular care to ensure we have not mistaken an activity cycle for a planetary signal.  In Figure \ref{197037bvs}, we show our measured BVS and $S_{HK}$ indices, which we use to evaluate the influence of stellar photospheric activity on the signal of HD 197037b.  Using both a Pearson correlation test and least-squares fitting, we find no correlation to the RVs for the BVS. While the $S_{HK}$ indices appear somewhat correlated with the RVs to the unaided eye, the Pearson correlation coefficient of -0.26 indicates that the relation is not statistically significant.  Periodogram analysis (Figure \ref{197037bvsps}) shows no power at the 1030 day peak in the BVS or $S_{HK}$ time series.  We do see a modestly (FAP = 0.01) significant peak at 19.1 days in the periodogram of our $S_{HK}$ measurements, which we speculate may be the stellar rotation period.  Although we are admittedly unable to completely rule out the possibility of stellar activity as the source of our observed RV signal, the lack of correlation between our activity indicators and the velocities, and the absence of periodicity in $S_{HK}$ and BVS around the fitted period lead us to the conclusion that a planetary orbit is the most likely cause.

While the stellar activity measurements reinforce the planetary nature of the primary RV signal of planet b, the analysis of the long-term trend is not so clear.  Our periodograms of the BVS and $S_{HK}$ series both show an increase in power at very long periods, matching the behavior of the trend in the residual RVs, albeit at much lower power.  Furthermore, we see a correlation between $S_{HK}$ and the residual RVS around the one-planet fit with a Pearson correlation coefficient of -0.33, which is significant for the size of our data set.  The residual RVs are uncorrelated with the BVS, though, so the trends may be coincidental.  HD 197037 will need continued monitoring to determine the true nature of this long-period signal.

\subsection{HD 220773}

\subsubsection{RV Data and Orbit Modeling}

Our RV data set (Table \ref{220773tab}) for HD 220773 consists of 43 HET/HRS spectra taken over 9 years between July 2002 and August 2011.  The data have an RMS of 11.7 m/s and a mean error of 4.11 m/s.

\begin{figure}
  \begin{center}
    \includegraphics[scale=0.6]{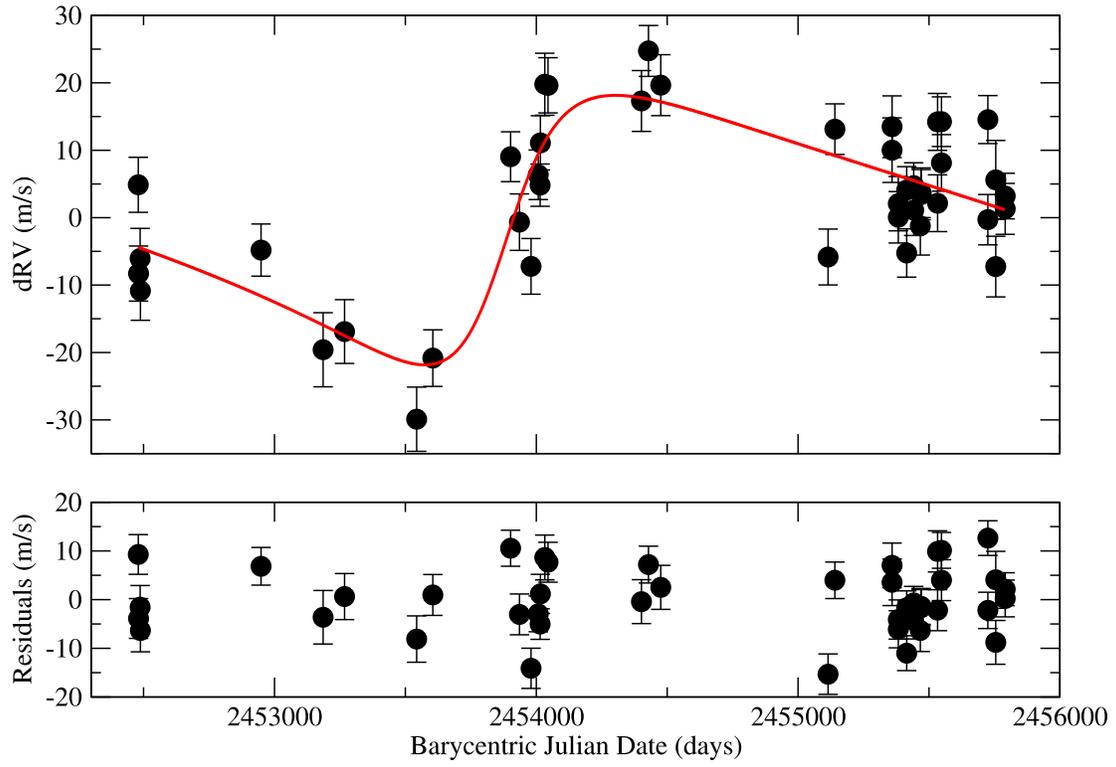}
    \caption{\emph{Top}: Radial velocity data for HD 220773.  The best-fit orbit model is shown as a red line.  \emph{Bottom}: Residuals to a one-planet fit.}
    \label{220773rv}
    \end{center}
\end{figure}

Based on our data, we find evidence for a highly eccentric giant planet on a long-period orbit.  Because of its high eccentricity, the period of planet b does not appear at significant power in our periodogram analysis.  However, the high RMS of our RV set, combined with the characteristic eccentric turnaround of the velocity time series (Figure \ref{220773rv}), strongly indicate the presence of a planetary companion.  In order to offset the lack of information from the periodogram, we have run GAUSSFIT with a broad range of periods (2500d-4500d) and eccentricities (0.4-0.7).  All fits converge unambiguously to a period of 3725 days with an eccentricity of 0.51, indicating a $1.45 M_J$ planet at 4.94 AU.  We include the plot of this model in Figure \ref{220773rv}.  The final fit gives a reduced $\chi^2$ of 3.14 and an RMS scatter of 6.57 m/s.  Because the reduced $\chi^2$ is so high, we have computed the model again after adding the 5.10 m/s ``jitter'' term in quadrature to the errors listed in Table \ref{220773tab}.  While our fitted parameters and uncertainties do not change significantly from the values given in Table \ref{orbits}, the reduced $\chi^2$ drops to 1.18, lending additional confidence to our solution.  We do not currently see any evidence for additional signals in the residual RVs.

\begin{figure}
  \begin{center}
    \includegraphics[scale=0.6, angle=-90]{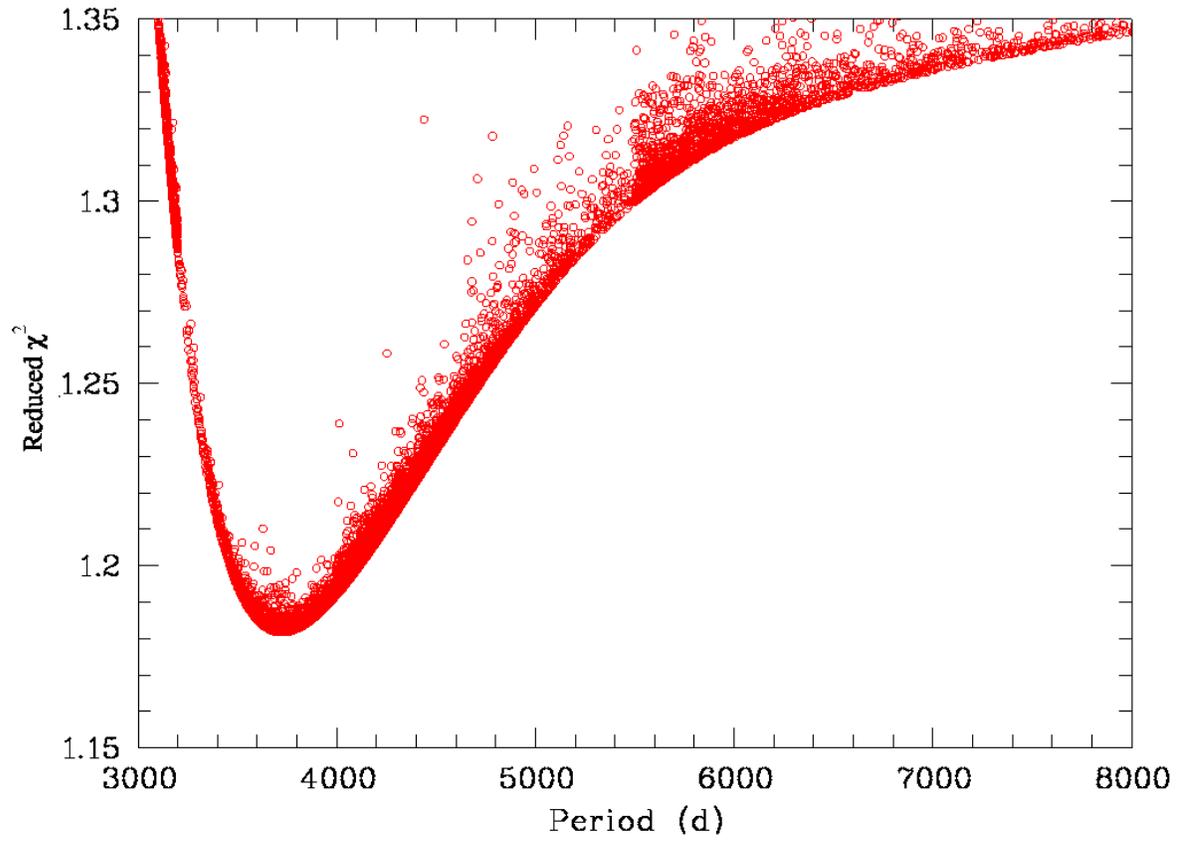}
    \caption{Genetic algorithm results for HD 220773.  Each circle indicates the minimum $\chi^2$ reached by fitting a one-planet orbit at the given period.}
    \label{220773ga}
   \end{center}
\end{figure}

To confirm the orbital parameters listed in Table \ref{orbits}, we use a genetic algorithm to explore the parameter space and evaluate the likelihood of the null hypothesis.  The algorithm fits a grid of parameters (number of planets, masses, periods, eccentricities) to the RV set, allowing for a thorough exploration of how $\chi^2$ behaves in response.  Areas of parameter space which do not match the data are iteratively rejected, allowing the routine to converge on an optimal solution.  We have performed 10,000 iterations of the algorithm with our RV data, considering periods between 3000 and 10,000 days.  For this experiment, we have again added the stellar ``jitter'' term to our measurement errors.  As shown in Figure \ref{220773ga}, the genetic algorithm reaches $\chi^2 < 1.20$ for a one-planet model with periods close to our fitted value in Table \ref{orbits}.  This is a dramatic improvement over a zero-planet model, which yields $\chi^2 = 3.65$.

\subsubsection{Stellar Activity and Line Bisector Analysis}

Our bisector velocity spans for HD 220773, which have an RMS scatter of 24 m/s, show no correlation to our measured RVs, and a periodogram analysis of the BVS shows no power around the period of planet b.  While the periodogram signal for the RVs did not meet the criterion for a positive detection, there was a broad peak centered around the $\sim 4000$ day period of the planet.  The BVS, on the other hand, show no evidence whatsoever for a long-period trend or signal.  We conclude, then, that the BVS do not indicate stellar activity which could mimic the behavior of this long-period planet.  Likewise, while the HRS does not provide time-series $S_{HK}$ information, our 2.7 m stellar spectrum does offer a ``snapshot'' value of $\log R'_{HK}$.  From this spectrum, we derive $\log R'_{HK} = -4.98$, which is consistent with the low activity level suggested by the BVS analysis.

\section{\bf Updated Planetary Parameters for HD 155358}

HD 155358 has previously been identified as a two-planet system by the McDonald Observatory Planet Search Program \citep{cochran07}.  It is a notable system because it hosts two giant planets despite having a measured [Fe/H] among the lowest of any stars with substellar companions of an unambiguously planetary nature (excepting HIP 13044 \citep{setiawan11}, which apparently originates outside the Galaxy).  \citet{fuhrmann08} analyze its chemical abundances in more detail, leading them to claim that the star is actually a member of the thick disk population of the Galaxy, a claim strengthened by the $>10$ Gyr age estimate of \citet{casagrande11}.  While our updated analysis suggests a metallicity of [Fe/H] = -0.51, rather than our original estimate of [Fe/H] = -0.68, the system is still extremely metal-poor relative to the other known exoplanetary systems.  Continued monitoring of HD 155358 has caused us to reevaluate our previously reported orbital parameters, resolving the ambiguity in the period of planet c discussed in \citet{cochran07}.

\subsection{RV Data and Orbit Modeling}

\begin{figure}
  \begin{center}
    \subfigure[\label{155358rv}]{\includegraphics[scale=0.45]{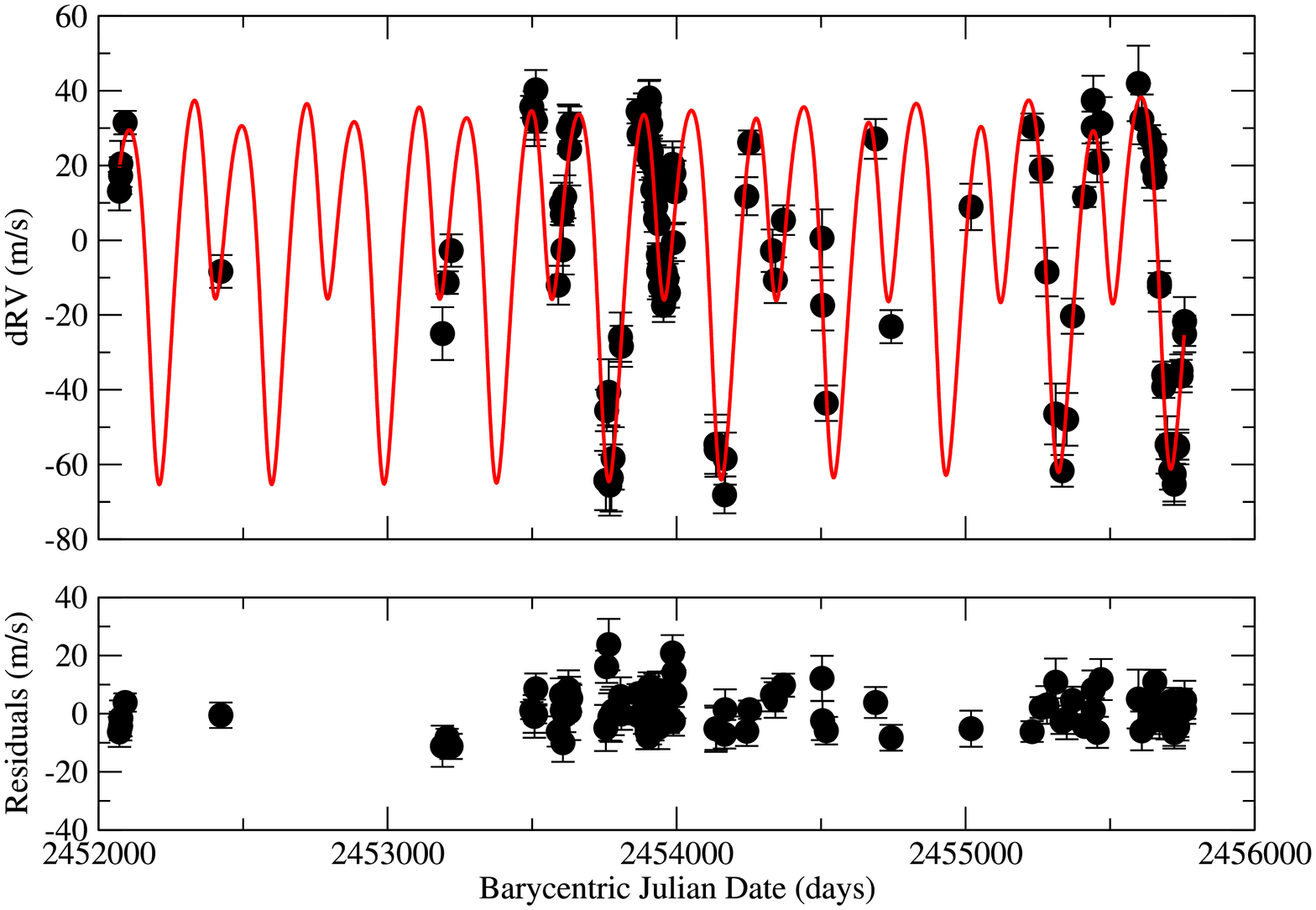}}
    \subfigure[\label{155358phase}]{\includegraphics[scale=0.45]{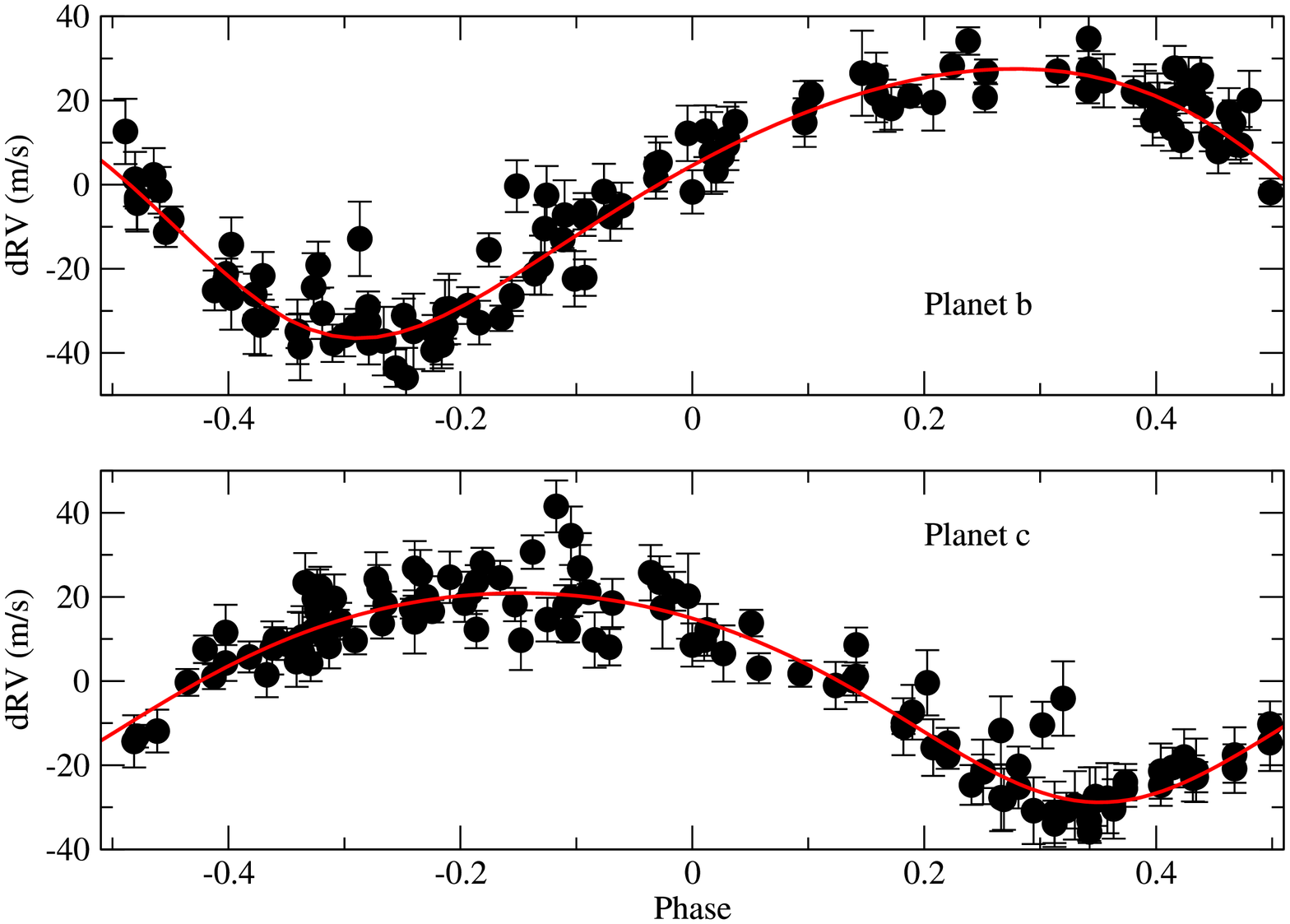}}
    \caption{a.  \emph{Top}: Radial velocity data for HD 155358.  The best-fit orbit model is shown as a red line.  \emph{Bottom}: Residuals to a two-planet fit.
\newline
\newline
b.  Phase plots for planets b (\emph{top}) and c (\emph{bottom}).  In each plot, the signal of the planet not shown has been subtracted from the RVs.}
    \end{center}
\end{figure}

Since the publication of \citet{cochran07}, we have obtained 51 additional RV points for HD 155358 from HET.  We have re-reduced all of our spectra with the latest version of AUSTRAL for consistency, and include all 122 points in Table \ref{155358tab}.  The entire set of velocities, shown as a time series in Figure \ref{155358rv}, covers approximately 10 years from June 2001 until August 2011.  These RVs have an RMS scatter of 30.2 m/s, much higher than we expect from a mean error of just 5.38 m/s.

\begin{figure}
  \begin{center}
    \subfigure[\label{155358ps}]{\includegraphics[scale=0.5]{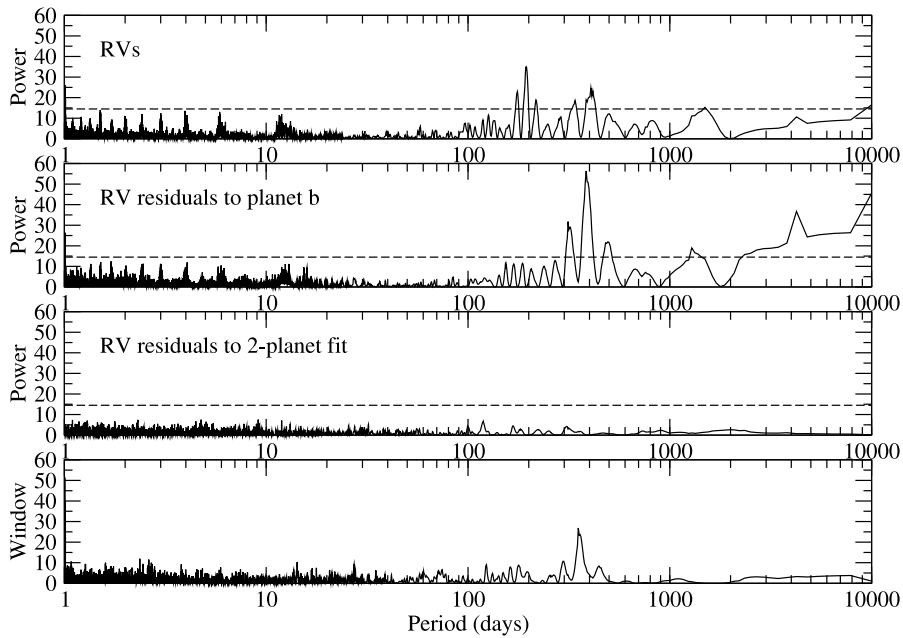}}
    \subfigure[\label{155358w}]{\includegraphics[scale=0.4]{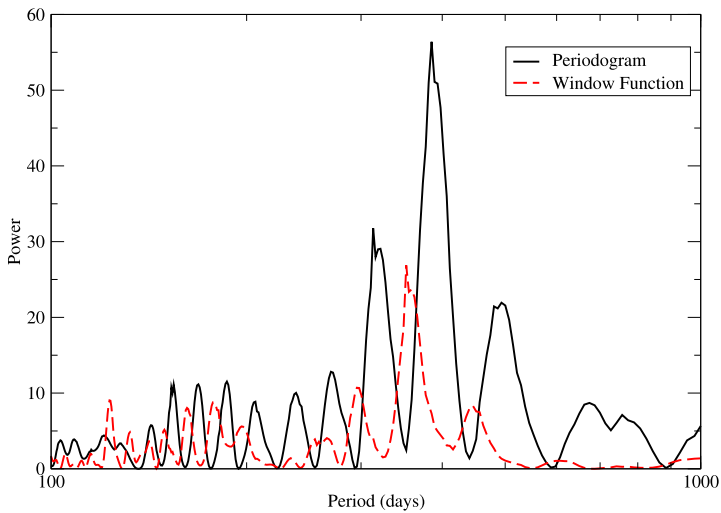}}
    \caption{a.  \emph{From top}: [\emph{1}] Generalized Lomb-Scargle periodogram for HD 155358 RVs.  [\emph{2}] The same periodogram for the residual RVs after subtracting a one-planet fit.  [\emph{3}] Periodogram of the residual RVs after subtracting a two-planet fit.  [\emph{4}] Periodogram of our time sampling (the window function).  The dashed lines indicate the approximate power level for a FAP of 0.01.
\newline
\newline
b.  Periodogram for planet c (black) with the window function (red).  Note that the two peaks do not overlap.}
    \end{center}
\end{figure}

We include the periodogram and window function for HD 155358 in Figure \ref{155358ps}.  As expected, we see a significant peak around 196 days, with a preliminary FAP of $1.15 \times 10^{-11}$.  Fitting a one-planet orbit from this peak, we find an RV amplitude $K = 32$ m/s with a period $P = 195$ days and an eccentricity of $e = 0.23$.  Based on this fit, HD 155358b has a minimum mass of 0.85 Jupiter masses at $a = 0.64$ AU.

\citet{cochran07} note that the periodogram for the residuals around the one-planet fit to planet b initially showed power around 530 days and 330 days.  Comparing fits at both periods produced better results at the longer period for our data at that time.  However, the periodogram of the residuals around planet b for our current data (Figure \ref{155358ps}) clearly indicates the shorter period as the true signal.  The peak at 391 days has a preliminary FAP too low for the precision of our code (approximately $10^{-14}$), and we no longer see additional peaks at longer periods.  While the window function for our sampling does show some power at the one-year alias, the period of this periodogram peak is sufficiently separated from the yearly alias (Figure \ref{155358w}) that we are confident the signal is not due to our sampling.

Using 391 days as a preliminary guess for the period of planet c, we performed a two-planet fit to our RVs.  This updated fit changes the parameters of planet b to $P = 194.3$ days, $M \sin i = 0.85 M_{J}$, and $e = 0.17$, with the orbital separation remaining at 0.64 AU.  Planet c then converges to a period of 391.9 days, with $e = 0.16$ and $K = 25$ m/s.  The derived properties for planet c then become $M = 0.82 M_J$ and $a = 1.02$ AU.  We plot this orbit over our RV data in Figure \ref{155358rv}.  The addition of planet c reduces our RMS to 6.14 m/s with a reduced $\chi^2$ of 1.41.  As a consistency check, we test a fit with parameters for planet c more closely matching those from \citet{cochran07}, but find no satisfactory solution at this longer period.  Holding the parameters for planet c fixed at the values derived in that study result in a reduced $\chi^2$ of 11.2 and a residual RMS of 15.4 m/s.

\subsection{Stellar Activity and Line Bisector Analysis}

The BVS for HD 155358 display an RMS scatter of 25 m/s, and are uncorrelated with both the RVs and the residuals to the 2-planet fit.  Additionally, we see no significant peaks in the periodogram of the BVS time series.  Also, while we present HET velocities here, we do have some 2.7 m spectra from a preliminary investigation of HD 155358, allowing us to examine the Ca H and K indices.  $S_{HK}$ shows very little activity, and we derive a $\log R'_{HK}$ of -4.54.  The hypothesis that the two large-amplitude signals observed here are due to stellar activity can therefore conclusively be ruled out.

\subsection{Dynamical Analysis of 3-body System}

Even at the separation claimed in \citet{cochran07}, planets b and c are close enough to interact gravitationally. Our updated orbital model now indicates they are actually much closer together. While their periods suggest the planets are in a 2:1 mean motion resonance, our preliminary orbital simulations showed the system to be unstable for a range of input values. We therefore decided to perform a highly detailed dynamical study of the system to investigate whether the orbits that best fit the data are indeed dynamically feasible. To do this, we performed over 100,000 unique simulations of the HD 155358 system using the Hybrid integrator within the $n$-body dynamics package MERCURY \citep{chambers99}.

To systematically address the stability of the HD 155358 system as a function of the orbits of planets b and c, we followed \citet{horner11} and \citet{marshall10}, and examined test systems in which the initial orbit of the planet with the most tightly constrained orbital parameters (in this case planet b) was held fixed at the nominal best fit values. The initial orbit of the outermost planet was then systematically changed from one simulation to the next, such that scenarios were tested for orbits spanning the full $\pm 3 \sigma$ error ranges in semi-major axis, eccentricity, longitude of periastron and mean anomaly. Such tests have already proven critical in confirming or rejecting planets suggested to move on unusual orbits \citep[e.g.][]{witt11b}, and allow the construction of detailed dynamical maps for the system in orbital element phase space.

Keeping the initial orbit of the innermost planet fixed, we examined 31 unique values of semi-major axis for planet c, ranging from 0.96 AU to 1.08 AU, inclusive, in even steps. For each of these 31 initial semi-major axes, we studied 31 values of orbital eccentricity, ranging from the smallest value possible (0.0) to a maximum of 0.46 (corresponding to the best fit value, 0.16, plus three sigma). For each of the 961 $a$-$e$ pairs, we considered 11 values of initial mean anomaly and initial longitude of periastron ($\omega$), resulting in a total suite of 116281 ($31 \times 31 \times 11 \times 11$) plausible architectures for the HD 155358 system.

In each simulation, the planet masses were set to their minimum ($M \sin i$) values; the mass of the innermost planet was set to $0.85 M_J$, and that of the outermost was set to $0.82 M_J$. As such, the dynamical stability maps obtained show the maximum stability possible (since increasing the masses of the planets would clearly increase the speed at which the system would destabilize for any nominally unstable architecture). The dynamical evolution of the two planets was then followed for a period of 100 million years, or until one of the planets either collided with the central star, was transferred to an orbit that took it to a distance of at least 10 AU from the central star, or collided with the other planet.

\begin{figure}
\begin{center}
  \subfigure[\label{meanae}]{\includegraphics[scale=0.8]{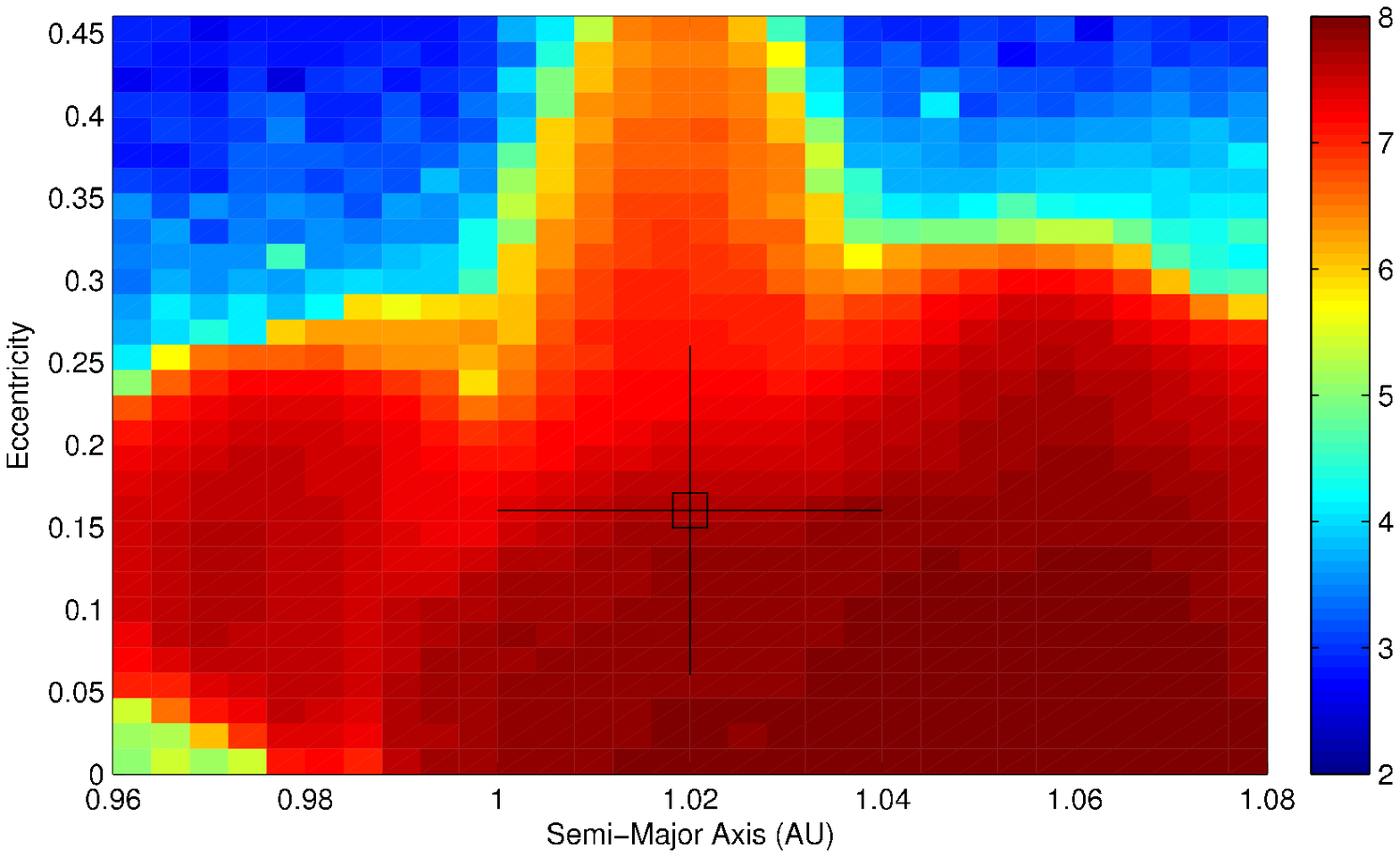}}
  \subfigure[\label{medae}]{\includegraphics[scale=0.8]{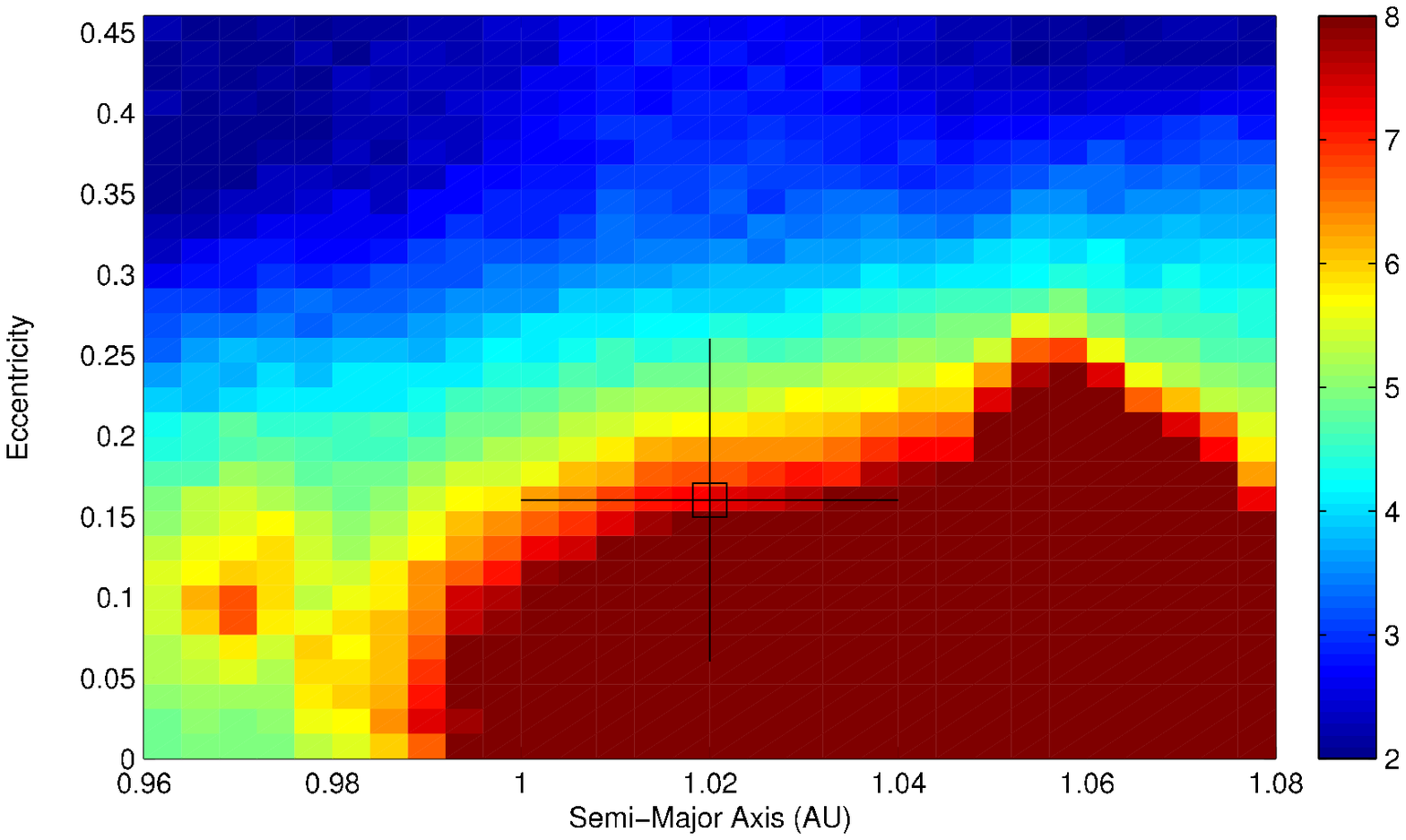}}
  \caption{Results of our stability analysis for the HD 155358 planetary system in $a-e$ space.  At each grid point, we have run 25 dynamic simulations with planet c at that point, with planet b remaining fixed at the configuration given in Table \ref{orbits}.  The resulting intensity indicates the mean (\emph{a}) or median (\emph{b}) lifetime of the planet over those 25 runs in log years.  Longer lifetimes indicate a more stable configuration.  The crosshairs indicate our best-fit parameters as derived from the RV data (Table \ref{orbits}).}
  \label{stabilitye}
\end{center}
\end{figure}

\begin{figure}
\begin{center}
  \subfigure[\label{meanaw}]{\includegraphics[scale=0.8]{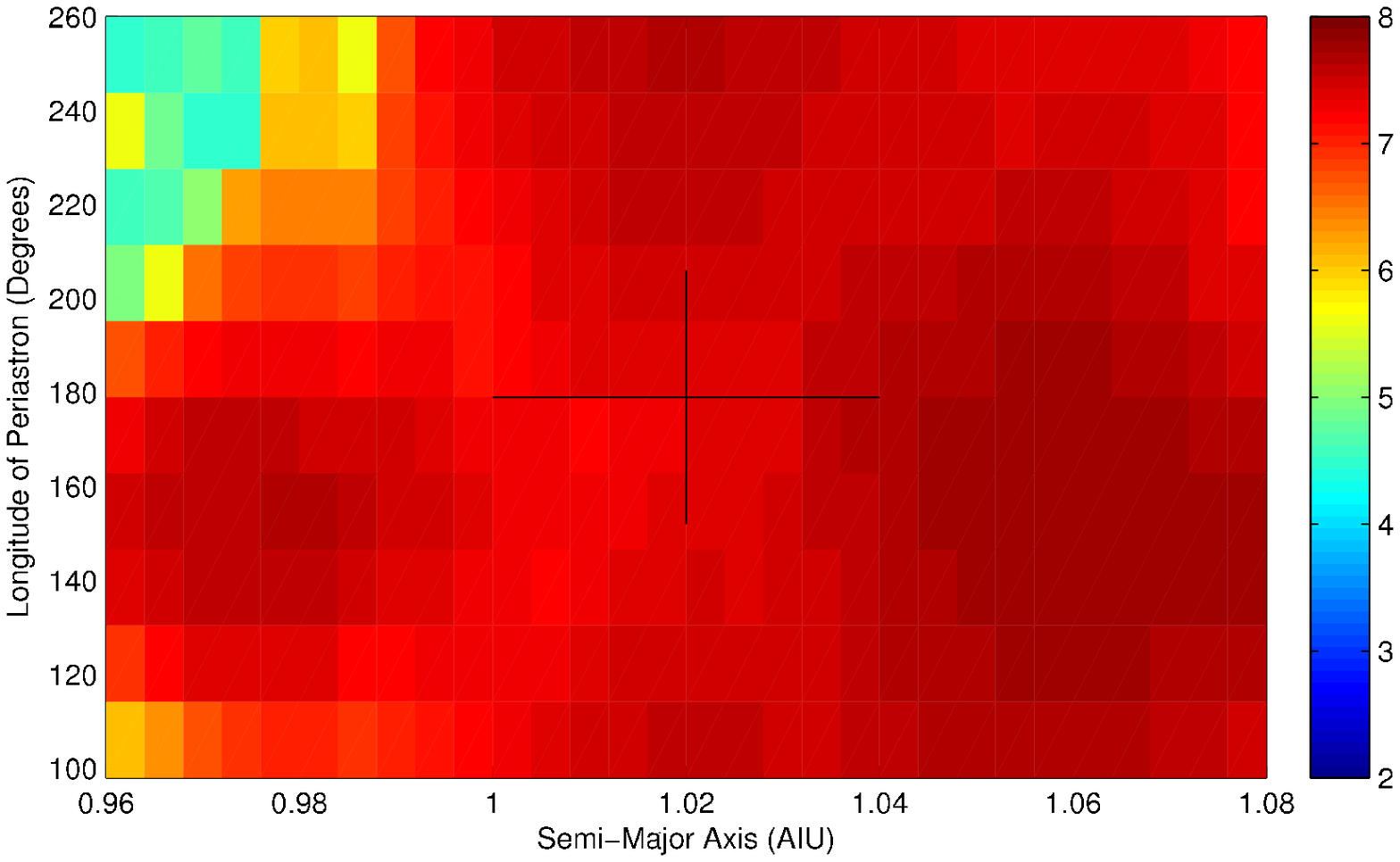}}
  \subfigure[\label{medaw}]{\includegraphics[scale=0.8]{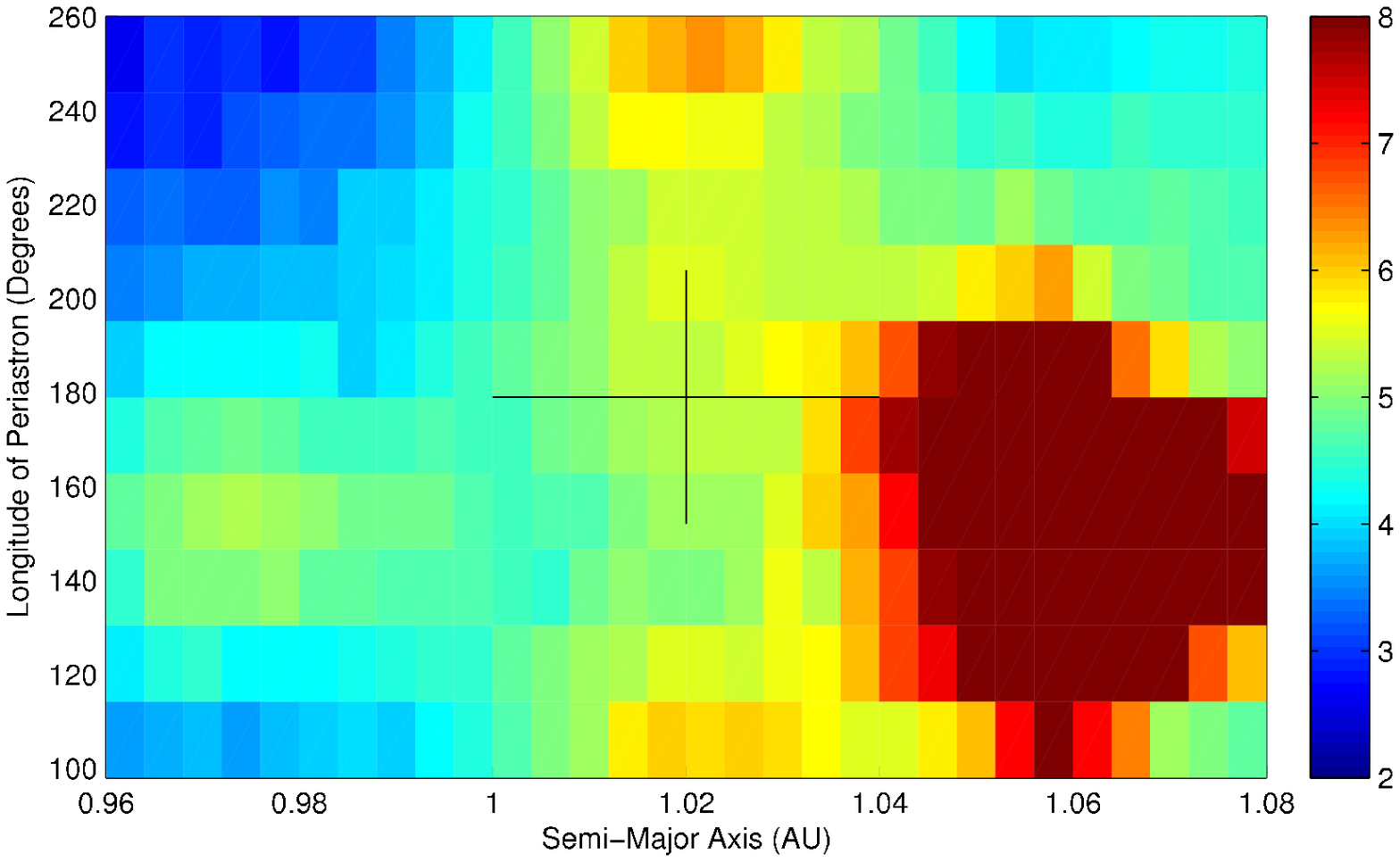}}
  \caption{Results of our stability analysis for the HD 155358 planetary system in $a-\omega$ space.  At each grid point, we have run 25 dynamic simulations with planet c at that point, with planet b remaining fixed at the configuration given in Table \ref{orbits}.  The resulting intensity indicates the mean (\emph{a}) or median (\emph{b}) lifetime of the planet over those 25 runs in log years.  Longer lifetimes indicate a more stable configuration.  The crosshairs indicate our best-fit parameters as derived from the RV data (Table \ref{orbits}).}
  \label{stabilityw}
\end{center}
\end{figure}

The results of our simulations are shown in Figures \ref{stabilitye}-\ref{stabilityw}. We present the multi-dimensional grid of orbital parameters over which we ran our simulations as two-dimensional cross sections, indicating the mean and median stability lifetimes over each of the runs at each grid point. So, for example, each grid point in our plots in $a$-$e$ space reveals either the mean or the median of 121 unique simulations of an orbit with those particular $a$-$e$ (i.e. 11 in $\omega$ times 11 in mean-anomaly). Similarly, each point in the $a$-$\omega$ plots corresponds to the mean (or median) of 341 separate trials (31 in $e$ times 11 in $\omega$).

The first thing that is apparent from Figure \ref{meanae}, which shows the mean lifetime of the system as a function of semi-major axis and eccentricity, is that the stability of the system is a strong function of eccentricity. Solutions with low orbital eccentricity are typically far more dynamically stable than those with high eccentricity. In addition, the stabilizing effect of the 2:1 mean-motion resonance between planets b and c can be clearly seen as offering a region of some stability to even high eccentricities between $a \sim 1.05$ and 1.25 AU. However, it is apparent when one examines the median lifetime plot (Figure \ref{medae}) that a significant fraction of eccentric orbits in that region are dynamically unstable (the reason for the apparently low median lifetimes in that region). The reason for this is that the dynamical stability of orbits in this region, particularly for high eccentricities, is also a strong function of the initial longitude of periastron for planet c's orbit. Indeed, comparison of this figure to those showing the influence of the longitude of periastron for planet c reveals that the most stable regions therein lie beyond the $1 \sigma$ error bars for the nominal orbit. For low eccentricities ($e < 0.1$), in the vicinity of the 2:1 resonance, then the orbit is stable regardless of the initial longitude of periastron, but for high eccentricities ($e > 0.15$), the stable regions lie towards the edge of the allowed parameter space, making such a solution seem relatively improbable.

It is apparent from close examination of Figure \ref{medae} (which shows the median stability as a function of semi-major axis and eccentricity) that the key determinant of the stability of the system (particularly for non-resonant orbits) is actually the periastron distance of planet c. The sculpted shape of the stability plot, outwards of $a \sim 0.98$ AU, is very similar to that observed for the proposed planets around the cataclysmic variable HU Aquarii \citep{horner11,witt11b}. As was found in that work, the dividing line between unstable and stable orbits again seems to fall approximately five Hill radii beyond the orbit of the innermost planet. Any orbit for planet c that approaches the orbit of b more closely than this distance will be unstable on astronomically short timescales (aside from those protected from close encounters by the effects of the 2:1 mean motion resonance between the planets). On the other hand, orbits which keep the two planets sufficiently far apart tend to be dynamically stable. We note, too, that the sharp inner cut-off to this broad region of stability, at around 0.99 AU, corresponds to the apastron distance of the innermost planet plus five Hill radii. Once again, orbital solutions that allow the two planets to approach closer than five times the Hill radius of the inner planet destabilize on relatively short timescales.

The results of our dynamical simulations show that a large range of dynamically stable solutions exist within the $1 \sigma$ errors on the orbit of HD 155358c. Given the breadth of the resonant feature apparent in Figure \ref{meanae} (the mean lifetime as a function of $a$ and $e$), it seems clear that all orbits within $1 \sigma$ of the nominal best fit will be strongly influenced by that broad resonance. As such, it seems fair to conclude that these two planets are most likely trapped within their mutual 2:1 mean-motion resonance, although it is also apparent that non-resonant solutions also exist that satisfy both the dynamical and observational constraints. The stability restrictions are consistent with our orbital fit, but the parameters listed in Table \ref{orbits} have not been modified to include the information obtained from the orbital simulations. 

\subsection{Habitability of Exomoons}

Using the [Fe/H] and T$_{eff}$ derived herein, an [$\alpha$/Fe] of 0.32 \citep{fuhrmann08}, and an age of 10.7 Gyr \citep{casagrande11}, we have fit Yonsei-Yale isochrones \citep{demarque04} to HD 155358.  With mass estimates ranging from $0.87 M_{\odot}$ \citep{lambert91} to $0.92 M_{\odot}$ \citep{casagrande11}, HD 155358 has a luminosity somewhere between $L_{\ast} = 1.14 L_{\odot}$ and $L_{\ast} = 1.67 L_{\odot}$.  If we assume the location of the habitable zone scales as $\sqrt{L_{\ast}}$, then at an orbital separation of 1.06 AU, planet c lies within the habitable zone of its parent star \citep[as defined by][]{kasting93}.  Typically, Jovian planets are not considered to be potential habitats for Earth-like life \citep{lammer09}. However, it has been suggested that such planets could host potentially habitable satellites \citep[e.g.][]{porter11}, or even that they could dynamically capture planet-sized objects during their migration as satellites or Trojan companions\footnote{Planetary Trojans are a particularly fascinating population of objects, trapped in 1:1 mean-motion resonance with their host planet. Typically, stable Trojans follow horseshoe-shaped paths that librate around either the leading- or trailing-Lagrange points in that planets orbit, located sixty degrees ahead and behind the planet \citep[for a good illustration of such orbits, we direct the interested reader to][]{horner06}. \\\\Within our Solar system, both Jupiter and Neptune host significant populations of Trojans that were captured during their migration \citep[e.g.][]{morbidelli05,lykawka09}. For those planets, the captured Trojans are typically small, but there is nothing to prevent a giant planet from capturing an Earth-mass object as a Trojan during its migration. Once captured, and once the migration stops, such objects can be dynamically stable on timescales of billions of years \citep[e.g.][]{hornerl10,lykawka11}, even when moving on orbits of significantly different eccentricity and inclination to their host planet. Whilst the detection of such planets would no doubt be challenging \citep[e.g.][]{ford07}, they remain an intriguing option in the search for habitable worlds.} \citep[e.g.][]{tinney11}. In fact, the presence of two giant planets on relatively close-in orbits may indirectly increase the water content on terrestrial satellites through radial mixing of planetesimals rich in ices \citep{mandell07}, a key requirement for such objects to be considered habitable \citep[e.g.][]{hornerj10}. Furthermore, any satellites sufficiently large to be considered habitable would also be subject to significant tidal heating from their host planet, which would likely act to increase their habitability when they lie towards the outer edge of Kasting's habitable zone (as they would have earlier in the life of the system). The induced tectonics would also potentially improve the habitability of any such moons \citep[e.g.][]{hornerj10}.

Our dynamical analysis suggests that the two planets in the HD 155358 system are most likely trapped in mutual resonance. It might initially seem that such orbits would be variable on the long term, such that they would experience sufficiently large excursions as to render any satellites or Trojan companions uninhabitable.  However, we know from our own Solar system that long term resonant captures can be maintained on timescales comparable to the age of the Solar system \citep[e.g. the Neptune Trojans,][]{lykawka10}.  As such, it is reasonable to assume that, if the planets are truly trapped in mutual resonance, they could have been on their current orbits for at least enough time for any moons or Trojans of planet c to be considered habitable. However, the current best-fit eccentricity for planet c is sufficiently high that it lies at the outer limit of the range in $e$ that would allow for a habitable exomoon \citep{tinney11}, which might limit the potential habitability of any exomoons in the system. That said, the one-sigma range of allowed orbits extends to relatively low eccentricities ($e=0.06$), which is certainly compatible with potential habitability. As such, it is possible that such moons, if present, could be potential habitats for life.

\section{\bf Discussion}

Of the planets discovered via the radial velocity method, only around 7 percent have $a \ge 3$ AU.  The addition of the planets presented here therefore represent a significant contribution to that sample.  As more such objects start to fall within the detection limits of RV surveys, our findings provide interesting comparison cases to begin to look for trends in the long-period gas giant population.

Continued monitoring of the HD 79498, HD 197037, and HD 220773 systems is crucial for refinement of planet formation and migration theories.  The presence or absence of additional smaller planets will shape our understanding of the migrational history of these systems.  \citet{mandell07}, for example, predict markedly different outcomes for the formation and water content of sub-giant planets after migration of a Jovian planet depending on whether or not gas drag plays a significant role in planetary migration.  

Of the four stars presented herein, at least one--HD 155358--hosts multiple gas giant planets.  Additionally, HD 197037 shows tentative evidence of a yet-undetected substellar companion at $a \sim 5.5$ AU.  If the occurrence of multiple-Jupiter systems is as common as it appears in this very small sample, it would support the claim by \citet{sumi11} that the large population of free-floating planetary-mass objects within the Galaxy form in protoplanetary disks as planets.  On the other hand, current RV results \citep{witt09} indicate that multi-giant planet systems should actually be quite rare, at least inside of 2-3 AU.  Understanding whether that result remains valid for planets further out, though, will rely heavily on detections of residual long-term trends for planets like HD 197037b.

The orbital evolution of the HD 155358 system is of particular interest, as it provides a comparison case for theoretical work exploring the formation and migration of the Jupiter-Saturn system \citep[e.g.][]{tsiganis05}.  While Jupiter and Saturn appear to have crossed the 2:1 MMR and later separated, the HD 155358 giants remain locked in resonance, presumably for the entire $\sim 10$ Gyr lifetime of the system.  Accounting for the difference in these final configurations most likely requires different initial architectures, disk masses, and encounter histories for the two systems.

While the stability simulations presented here offer some constraints on the orbital configurations of the HD 155358 system, the true geometries of those planets' orbits will only be fully understood if their mutual inclinations are measured. Unfortunately, the HD 155358 planets will only be accessible to the next generation of astrometric instrumentation, as their predicted astrometric displacements are below the 0.2 mas precision limit of the HST Fine Guidance Sensor \citep{nelan10}.  On the other hand, HD 220773b, which has a predicted displacement of 0.242 mas, might be an interesting astrometry target for FGS, both for the purpose of determining a true mass, and to search for outer companions to this distant planet.

The refined orbital parameters for HD 155358c places it, to zeroth order, within the habitable zone.  While the planet itself is likely inhospitable to any Earthlike lifeforms, the possibility of habitable moons or Trojans make it an interesting datum for examining more exotic environments for biology \citep{dsm11}.  Given the advanced age of the system, any intelligent life residing on planet c is potentially far more advanced than our own civilization.  The technological advances accompanying such extended development may make interstellar broadcasts or beacons energetically and financially feasible, making HD 155358 an interesting target for SETI \citep[e.g.][]{benfordj10,benfordg10}

Our results are somewhat atypical when viewed in the context of the metallicity-frequency correlation for giant planets \citep[see, e.g.][]{fischer05}.  Only HD 79498 has significantly super-solar metallicity, while HD 197037 and HD 155358 qualify as metal-poor.  In the case of HD 197037, its higher mass and earlier spectral type increase its likelihood of forming Jovian planets, at least somewhat offsetting any metallicity effects \citep{johnson11}.  HD 155358, on the other hand, remains a true anomaly despite the slightly higher [Fe/H] reported herein.  While the sample presented here is obviously too small to make even tentative statements about the validity of observed correlations, it will be interesting to see whether gas giants continue to be found preferentially around metal-rich stars as long-term RV surveys begin to reveal a large number of Jupiter-mass planets.  If the removal of the period bias on the giant planet census reveals a large population of Jovian planets around metal-poor stars, it will serve as strong evidence that at least some gas giants form through gravitational instability of the protoplanetary disk rather than core accretion \citep{boss02}.  Furthermore, the discovery of thick disk planets suggests a planet formation history essentially spanning the age of the Galaxy.

\begin{acknowledgements}
The authors wish to thank Anita Cochran, Candace Gray, and Diane Paulson for their contributions to the enormous observational effort that went into this study.  M.E. and W.D.C. acknowledge support by the National Aeronautics and Space Administration under Grants NNX07AL70G and NNX09AB30G issued through the Origins of Solar Systems Program.  This research has made use of the Exoplanet Orbit Database and the Exoplanet Data Explorer at exoplanets.org.
\end{acknowledgements}

\clearpage

\clearpage

\begin{center}

\footnotesize

\tablecaption{Radial Velocities for HD 79498}
\label{79498tab}	
\tablefirsthead{\hline
BJD - 2450000 & Radial Velocity & Uncertainty & $S_{HK}$ \\
& (m/s) & (m/s) & \\
\hline}
		
\tablehead{\hline
\emph{Table \ref{79498tab} cont'd.} & & & \\ \hline
BJD - 2450000 & Radial Velocity & Uncertainty & $S_{HK}$ \\
& (m/s) & (m/s) & \\
\hline}
		
\tabletail{\hline}

\begin{supertabular}{| l l l l |}

3038.86139200 	 & -14.42 	 & 8.32 	 & $ 0.1823 \pm 0.0247 $ \\
3066.79412600 	 & -24.59 	 & 4.67 	 & $ 0.1709 \pm 0.0177 $ \\
3073.81710000 	 & -21.56 	 & 5.65 	 & $ 0.1598 \pm 0.0202 $ \\
3746.91651500 	 & 27.61 	 & 4.58 	 & $ 0.1532 \pm 0.0242 $ \\
3787.81423800 	 & 19.33 	 & 4.33 	 & $ 0.1592 \pm 0.0195 $ \\
3806.70783400 	 & 15.78 	 & 2.43 	 & $ 0.1584 \pm 0.0203 $ \\
3840.75580400 	 & 28.03 	 & 5.25 	 & $ 0.1848 \pm 0.0238 $ \\
3861.68232900 	 & 29.18 	 & 6.57 	 & $ 0.1535 \pm 0.0187 $ \\
4067.98581900 	 & 7.79 	 & 3.44 	 & $ 0.1435 \pm 0.0210 $ \\
4068.90695600 	 & 23.21 	 & 4.53 	 & $ 0.1607 \pm 0.0214 $ \\
4191.80349000 	 & 23.76 	 & 3.77 	 & $ 0.1617 \pm 0.0220 $ \\
4569.73532700 	 & 9.86 	 & 3.82 	 & $ 0.1748 \pm 0.0216 $ \\
4569.74801300 	 & 5.54 	 & 3.92 	 & $ 0.1701 \pm 0.0212 $ \\
4605.64036800 	 & 5.89 	 & 3.41 	 & $ 0.1759 \pm 0.0235 $ \\
4634.62414000 	 & 4.64 	 & 5.12 	 & $ 0.1701 \pm 0.0209 $ \\
4782.01348700 	 & 0.79 	 & 3.11 	 & $ 0.1590 \pm 0.0212 $ \\
4783.98080700 	 & -4.54 	 & 4.65 	 & $ 0.1640 \pm 0.0230 $ \\
4821.93818500 	 & -3.04 	 & 5.12 	 & $ 0.1705 \pm 0.0249 $ \\
4840.97668200 	 & -10.65 	 & 4.49 	 & $ 0.1599 \pm 0.0237 $ \\
4868.91890100 	 & -10.63 	 & 4.53 	 & $ 0.1651 \pm 0.0212 $ \\
4868.93042200 	 & -5.90 	 & 4.35 	 & $ 0.1710 \pm 0.0226 $ \\
4868.94194000 	 & -10.23 	 & 3.62 	 & $ 0.1693 \pm 0.0225 $ \\
4906.70221500 	 & -9.93 	 & 3.16 	 & $ 0.1663 \pm 0.0204 $ \\
4906.71734700 	 & -10.49 	 & 4.07 	 & $ 0.1578 \pm 0.0197 $ \\
4907.62242900 	 & -4.23 	 & 4.12 	 & $ 0.1602 \pm 0.0187 $ \\
4907.63742600 	 & -13.71 	 & 3.10 	 & $ 0.1629 \pm 0.0187 $ \\
4928.68528400 	 & -11.02 	 & 4.05 	 & $ 0.1802 \pm 0.0204 $ \\
4928.70027200 	 & -15.00 	 & 4.95 	 & $ 0.1845 \pm 0.0191 $ \\
4928.71526200 	 & -7.46 	 & 4.88 	 & $ 0.1761 \pm 0.0204 $ \\
4931.63355900 	 & -18.23 	 & 6.43 	 & $ 0.1698 \pm 0.0194 $ \\
4931.64625200 	 & -14.05 	 & 4.44 	 & $ 0.1657 \pm 0.0190 $ \\
4931.65894600 	 & -6.45 	 & 3.87 	 & $ 0.1693 \pm 0.0210 $ \\
4964.66292000 	 & -16.12 	 & 5.03 	 & $ 0.1610 \pm 0.0186 $ \\
4986.64365700 	 & -16.87 	 & 3.78 	 & $ 0.1659 \pm 0.0199 $ \\
5137.98754800 	 & -33.34 	 & 5.93 	 & $ 0.1727 \pm 0.0252 $ \\
5152.96977000 	 & -23.72 	 & 4.56 	 & $ 0.1580 \pm 0.0212 $ \\
5152.98476800 	 & -28.36 	 & 1.92 	 & $ 0.1594 \pm 0.0212 $ \\
5152.99976400 	 & -34.83 	 & 3.19 	 & $ 0.1755 \pm 0.0260 $ \\
5153.98375300 	 & -27.49 	 & 2.26 	 & $ 0.1751 \pm 0.0235 $ \\
5153.99875100 	 & -24.33 	 & 6.65 	 & $ 0.1744 \pm 0.0227 $ \\
5154.01374700 	 & -27.91 	 & 4.50 	 & $ 0.1941 \pm 0.0244 $ \\
5173.01441600 	 & -28.77 	 & 3.36 	 & $ 0.1637 \pm 0.0215 $ \\
5222.85632500 	 & -19.26 	 & 5.33 	 & $ 0.1604 \pm 0.0231 $ \\
5222.87135400 	 & -9.87 	 & 5.73 	 & $ 0.1657 \pm 0.0226 $ \\
5254.79363200 	 & -4.05 	 & 4.53 	 & $ 0.1638 \pm 0.0211 $ \\
5254.80862500 	 & -9.12 	 & 3.76 	 & $ 0.1625 \pm 0.0211 $ \\
5254.82361900 	 & 1.41 	 & 1.30 	 & $ 0.1625 \pm 0.0202 $ \\
5278.74248800 	 & 9.95 	 & 3.47 	 & $ 0.1674 \pm 0.0233 $ \\
5290.64067500 	 & 7.71 	 & 3.50 	 & $ 0.1597 \pm 0.0182 $ \\
5311.71658900 	 & 24.55 	 & 5.50 	 & $ 0.2135 \pm 0.0233 $ \\
5314.69970700 	 & 10.85 	 & 4.08 	 & $ 0.2051 \pm 0.0241 $ \\
5347.65756900 	 & 23.91 	 & 3.41 	 & $ 0.1536 \pm 0.0215 $ \\
5468.97994600 	 & 12.33 	 & 3.61 	 & $ 0.1438 \pm 0.0283 $ \\
5493.95838700 	 & 18.82 	 & 3.93 	 & $ 0.1603 \pm 0.0239 $ \\
5523.92790400 	 & 16.50 	 & 4.50 	 & $ 0.1629 \pm 0.0231 $ \\
5527.90018100 	 & 16.64 	 & 3.61 	 & $ 0.1651 \pm 0.0233 $ \\
5528.98017000 	 & 14.76 	 & 2.61 	 & $ 0.1598 \pm 0.0214 $ \\
5548.88450200 	 & 24.70 	 & 3.61 	 & $ 0.1526 \pm 0.0235 $ \\
5583.86404700 	 & 21.17 	 & 2.35 	 & $ 0.1635 \pm 0.0246 $ \\
5615.70061600 	 & 7.11 	 & 4.07 	 & $ 0.1584 \pm 0.0236 $ \\
5632.72839900 	 & 20.16 	 & 4.44 	 & $ 0.1402 \pm 0.0184 $ \\
5645.82335600 	 & 15.88 	 & 4.48 	 & $ 0.1827 \pm 0.0276 $ \\
5667.76838200 	 & 24.85 	 & 2.11 	 & $ 0.1616 \pm 0.0218 $ \\
5671.67914300 	 & 22.60 	 & 3.44 	 & $ 0.1665 \pm 0.0212 $ \\
5699.62389400 	 & 24.99 	 & 2.69 	 & $ 0.1654 \pm 0.0227 $ \\

\end{supertabular}

\end{center}

\begin{center}

\footnotesize

\tablecaption{Radial Velocities for HD 155358}
\label{155358tab}	
\tablefirsthead{\hline
BJD - 2450000 & Radial Velocity & Uncertainty \\
& (m/s) & (m/s) \\
\hline}
		
\tablehead{\hline
\emph{Table \ref{155358tab} cont'd.} & & \\ \hline
BJD - 2450000 & Radial Velocity & Uncertainty \\
& (m/s) & (m/s) \\
\hline}
		
\tabletail{\hline}

\begin{supertabular}{| l l l |}

2071.90483700 	 & 15.90 	 & 5.14 \\
2075.88682700 	 & 20.05 	 & 4.94 \\
2076.88985600 	 & 23.21 	 & 6.16 \\
2091.84643500 	 & 34.24 	 & 3.14 \\
2422.93714300 	 & -5.59 	 & 4.38 \\
3189.83913000 	 & -22.21 	 & 7.08 \\
3205.79562900 	 & -8.55 	 & 3.01 \\
3219.75321500 	 & 0.01 	 & 4.32 \\
3498.77697700 	 & 38.45 	 & 2.94 \\
3507.95763400 	 & 35.30 	 & 7.35 \\
3507.96287000 	 & 35.35 	 & 5.72 \\
3511.95319500 	 & 34.61 	 & 3.07 \\
3512.93931400 	 & 43.01 	 & 5.29 \\
3590.73322900 	 & -9.26 	 & 5.19 \\
3601.70090300 	 & 12.52 	 & 5.64 \\
3604.70193200 	 & 9.70 	 & 2.00 \\
3606.70299700 	 & 0.19 	 & 6.58 \\
3612.68403800 	 & 14.40 	 & 5.74 \\
3625.64546400 	 & 32.49 	 & 6.59 \\
3628.62430900 	 & 32.53 	 & 6.13 \\
3629.61946400 	 & 27.18 	 & 9.73 \\
3633.61627500 	 & 33.94 	 & 4.62 \\
3755.04213400 	 & -61.50 	 & 7.92 \\
3758.04274400 	 & -42.78 	 & 5.55 \\
3765.03555800 	 & -37.91 	 & 8.84 \\
3769.02604800 	 & -62.77 	 & 8.14 \\
3774.02174300 	 & -60.84 	 & 8.97 \\
3779.97987100 	 & -55.59 	 & 8.35 \\
3805.93069400 	 & -23.17 	 & 6.60 \\
3808.89282200 	 & -25.61 	 & 5.47 \\
3866.96427700 	 & 37.27 	 & 3.26 \\
3869.95498800 	 & 31.11 	 & 2.93 \\
3881.94511600 	 & 35.50 	 & 3.67 \\
3889.68235900 	 & 35.73 	 & 6.32 \\
3894.67240300 	 & 34.58 	 & 3.73 \\
3897.89540600 	 & 28.71 	 & 5.97 \\
3898.68534400 	 & 33.44 	 & 6.64 \\
3899.89225300 	 & 29.64 	 & 4.49 \\
3902.65793200 	 & 25.22 	 & 4.23 \\
3903.90645200 	 & 39.66 	 & 5.72 \\
3904.66233800 	 & 36.68 	 & 3.76 \\
3905.86023400 	 & 40.83 	 & 4.86 \\
3907.62941500 	 & 27.22 	 & 3.36 \\
3908.87691000 	 & 24.11 	 & 5.19 \\
3910.63980200 	 & 33.87 	 & 5.72 \\
3911.84525200 	 & 26.41 	 & 3.55 \\
3912.63613800 	 & 26.48 	 & 4.33 \\
3917.64144300 	 & 16.28 	 & 3.31 \\
3924.81885900 	 & 21.80 	 & 6.37 \\
3925.80799400 	 & 18.30 	 & 5.54 \\
3926.82465700 	 & 8.51 	 & 3.52 \\
3927.82203900 	 & 11.81 	 & 2.97 \\
3936.79096100 	 & -1.16 	 & 2.34 \\
3937.79035200 	 & -5.54 	 & 7.51 \\
3937.80465200 	 & 7.17 	 & 6.49 \\
3941.76239700 	 & -4.15 	 & 4.90 \\
3943.76691400 	 & -9.60 	 & 2.65 \\
3954.75689500 	 & -14.66 	 & 4.44 \\
3956.73476700 	 & -12.62 	 & 5.02 \\
3958.74755200 	 & -10.15 	 & 3.92 \\
3960.71484300 	 & -5.65 	 & 3.64 \\
3966.70814000 	 & -7.53 	 & 4.04 \\
3971.69798700 	 & -11.28 	 & 3.98 \\
3985.65146600 	 & 23.07 	 & 6.17 \\
3988.63745600 	 & 2.12 	 & 4.94 \\
3990.63945400 	 & 20.65 	 & 6.94 \\
3993.64492500 	 & 15.75 	 & 8.30 \\
4136.01098700 	 & -51.82 	 & 7.94 \\
4137.00571500 	 & -53.22 	 & 7.28 \\
4165.93375900 	 & -65.36 	 & 4.93 \\
4167.92406700 	 & -55.67 	 & 6.97 \\
4242.72348100 	 & 14.56 	 & 5.08 \\
4252.92738200 	 & 28.95 	 & 3.17 \\
4331.70165200 	 & -0.03 	 & 5.67 \\
4341.67683000 	 & -7.91 	 & 6.11 \\
4369.61238300 	 & 8.15 	 & 3.97 \\
4503.00826800 	 & 3.28 	 & 7.75 \\
4505.00883000 	 & -14.65 	 & 6.71 \\
4517.96440000 	 & -40.82 	 & 4.73 \\
4688.73045100 	 & 29.89 	 & 5.32 \\
4742.59078800 	 & -20.34 	 & 4.44 \\
5018.84957500 	 & 11.69 	 & 6.21 \\
5230.01528700 	 & 33.11 	 & 3.56 \\
5261.94061600 	 & 21.80 	 & 3.57 \\
5281.88078900 	 & -5.74 	 & 6.49 \\
5311.81103300 	 & -43.71 	 & 8.13 \\
5333.96546600 	 & -58.91 	 & 4.25 \\
5349.91269100 	 & -45.15 	 & 7.05 \\
5369.85529100 	 & -17.54 	 & 4.71 \\
5411.75132000 	 & 14.37 	 & 2.68 \\
5441.66382200 	 & 32.94 	 & 4.27 \\
5441.68951000 	 & 40.21 	 & 6.56 \\
5455.63610500 	 & 23.60 	 & 5.33 \\
5468.60493300 	 & 34.04 	 & 7.05 \\
5598.01915300 	 & 44.73 	 & 10.12 \\
5609.96969700 	 & 35.11 	 & 6.67 \\
5635.90270500 	 & 30.45 	 & 3.11 \\
5647.87224300 	 & 22.39 	 & 5.60 \\
5654.85070000 	 & 19.53 	 & 6.12 \\
5654.85347500 	 & 26.99 	 & 4.15 \\
5670.79410300 	 & -9.55 	 & 6.76 \\
5670.79685700 	 & -8.67 	 & 2.64 \\
5685.78224300 	 & -36.44 	 & 2.97 \\
5685.78500500 	 & -33.31 	 & 3.62 \\
5697.73042500 	 & -51.99 	 & 7.66 \\
5697.73318400 	 & -51.85 	 & 3.99 \\
5709.69993000 	 & -58.87 	 & 5.08 \\
5709.70269600 	 & -53.93 	 & 4.69 \\
5721.89536400 	 & -59.79 	 & 7.32 \\
5721.89813000 	 & -62.53 	 & 5.51 \\
5733.64282800 	 & -52.33 	 & 3.59 \\
5733.64560600 	 & -52.53 	 & 4.65 \\
5745.83215500 	 & -33.58 	 & 4.35 \\
5745.83491700 	 & -32.08 	 & 4.32 \\
5757.80163500 	 & -22.31 	 & 4.89 \\
5757.80440100 	 & -18.97 	 & 6.54 \\
5769.76209100 	 & -9.53 	 & 7.54 \\
5769.76486400 	 & -11.16 	 & 3.59 \\
5782.71759900 	 & 2.27 	 & 6.58 \\
5782.72037000 	 & -0.97 	 & 5.78 \\
5794.70209000 	 & 10.86 	 & 6.73 \\
5794.70486500 	 & 15.27 	 & 5.42 \\

\end{supertabular}

\end{center}

\clearpage

\begin{center}

\footnotesize

\tablecaption{Radial Velocities for HD 197037}
\label{197037tab}	
\tablefirsthead{\hline
BJD - 2450000 & Radial Velocity & Uncertainty & $S_{HK}$ \\
& (m/s) & (m/s) & \\
\hline}
		
\tablehead{\hline
\emph{Table \ref{197037tab} cont'd.} & & & \\ \hline
BJD - 2450000 & Radial Velocity & Uncertainty & $S_{HK}$ \\
& (m/s) & (m/s) & \\
\hline}
		
\tabletail{\hline}

\begin{supertabular}{| l l l l |}

1918.55952342 	 & 4.60 	 & 6.99 	 & $ 0.1506 \pm 0.0207 $ \\
2117.81076505 	 & -14.85 	 & 6.37 	 & $ 0.1610 \pm 0.0234 $ \\
2141.79847525 	 & -15.01 	 & 6.05 	 & $ 0.1629 \pm 0.0244 $ \\
2218.67152278 	 & -5.63 	 & 6.26 	 & $ 0.1520 \pm 0.0210 $ \\
2451.93562208 	 & 26.73 	 & 5.60 	 & $ 0.1611 \pm 0.0231 $ \\
2471.88269604 	 & 18.07 	 & 6.37 	 & $ 0.1591 \pm 0.0228 $ \\
2493.79718560 	 & 20.89 	 & 7.15 	 & $ 0.1648 \pm 0.0231 $ \\
2538.74935239 	 & 27.04 	 & 8.62 	 & $ 0.1604 \pm 0.0214 $ \\
2576.61483724 	 & 23.72 	 & 6.90 	 & $ 0.1579 \pm 0.0215 $ \\
2598.60530888 	 & 24.87 	 & 6.41 	 & $ 0.1656 \pm 0.0207 $ \\
2600.63887515 	 & 27.60 	 & 6.16 	 & $ 0.1606 \pm 0.0206 $ \\
2619.60435179 	 & 14.33 	 & 6.20 	 & $ 0.1556 \pm 0.0199 $ \\
2620.57904934 	 & 25.72 	 & 6.30 	 & $ 0.1646 \pm 0.0201 $ \\
2805.90471501 	 & 10.70 	 & 6.85 	 & $ 0.1802 \pm 0.0265 $ \\
2839.84681367 	 & -2.86 	 & 6.40 	 & $ 0.1646 \pm 0.0242 $ \\
2895.80593278 	 & -5.91 	 & 6.95 	 & $ 0.1617 \pm 0.0230 $ \\
2930.69425024 	 & 4.31 	 & 6.31 	 & $ 0.1652 \pm 0.0223 $ \\
2960.60916112 	 & -7.12 	 & 7.56 	 & $ 0.1678 \pm 0.0204 $ \\
3320.59580442 	 & -13.35 	 & 10.17 	 & $ 0.1804 \pm 0.0211 $ \\
3563.95949295 	 & 7.98 	 & 8.18 	 & $ 0.1593 \pm 0.0230 $ \\
3630.71472969 	 & 1.69 	 & 8.58 	 & $ 0.1827 \pm 0.0248 $ \\
3630.87333165 	 & -3.65 	 & 14.37 	 & $ 0.1866 \pm 0.0227 $ \\
3632.68024096 	 & 16.56 	 & 6.77 	 & $ 0.1641 \pm 0.0229 $ \\
3632.83643682 	 & -2.79 	 & 8.61 	 & $ 0.1543 \pm 0.0223 $ \\
3633.63266139 	 & 18.63 	 & 7.49 	 & $ 0.1631 \pm 0.0219 $ \\
3633.81885619 	 & 15.61 	 & 7.83 	 & $ 0.1592 \pm 0.0208 $ \\
3634.65565355 	 & 15.49 	 & 7.93 	 & $ 0.1635 \pm 0.0221 $ \\
3635.68024320 	 & 14.26 	 & 7.62 	 & $ 0.1678 \pm 0.0235 $ \\
3636.66099343 	 & 23.03 	 & 8.64 	 & $ 0.1667 \pm 0.0240 $ \\
3655.70765286 	 & 21.04 	 & 7.03 	 & $ 0.1592 \pm 0.0222 $ \\
3690.65885729 	 & -9.05 	 & 9.38 	 & $ 0.2181 \pm 0.0265 $ \\
3906.92067972 	 & -7.81 	 & 6.56 	 & $ 0.1677 \pm 0.0234 $ \\
3930.81642848 	 & -21.98 	 & 6.84 	 & $ 0.1657 \pm 0.0234 $ \\
3969.90052141 	 & -9.32 	 & 7.53 	 & $ 0.1732 \pm 0.0228 $ \\
3985.77426805 	 & -6.83 	 & 6.93 	 & $ 0.1674 \pm 0.0231 $ \\
4019.63256032 	 & -4.18 	 & 8.86 	 & $ 0.1651 \pm 0.0224 $ \\
4020.80723692 	 & -6.45 	 & 6.65 	 & $ 0.1603 \pm 0.0215 $ \\
4068.66375417 	 & 9.06 	 & 6.96 	 & $ 0.1601 \pm 0.0203 $ \\
4309.85063550 	 & -19.35 	 & 6.90 	 & $ 0.1770 \pm 0.0252 $ \\
4344.70987089 	 & 1.01 	 & 7.57 	 & $ 0.1680 \pm 0.0244 $ \\
4344.72024432 	 & -23.29 	 & 7.32 	 & $ 0.1631 \pm 0.0242 $ \\
4344.73061543 	 & -15.13 	 & 6.33 	 & $ 0.1695 \pm 0.0251 $ \\
4345.71065534 	 & -4.07 	 & 7.42 	 & $ 0.1728 \pm 0.0244 $ \\
4345.71987193 	 & -11.29 	 & 7.25 	 & $ 0.1677 \pm 0.0240 $ \\
4345.72908610 	 & -3.99 	 & 7.69 	 & $ 0.1601 \pm 0.0213 $ \\
4346.70200957 	 & -6.59 	 & 8.41 	 & $ 0.1670 \pm 0.0230 $ \\
4346.71122130 	 & 1.56 	 & 7.33 	 & $ 0.1670 \pm 0.0224 $ \\
4346.72043545 	 & 0.46 	 & 7.49 	 & $ 0.1655 \pm 0.0229 $ \\
4375.64708486 	 & -11.15 	 & 7.87 	 & $ 0.1714 \pm 0.0230 $ \\
4375.65745814 	 & -10.85 	 & 12.32 	 & $ 0.2151 \pm 0.0265 $ \\
4376.66244438 	 & -8.84 	 & 7.66 	 & $ 0.1685 \pm 0.0235 $ \\
4376.67281521 	 & -7.40 	 & 7.64 	 & $ 0.1679 \pm 0.0234 $ \\
4376.68318859 	 & -11.25 	 & 7.20 	 & $ 0.1701 \pm 0.0250 $ \\
4401.57166865 	 & -3.54 	 & 7.43 	 & $ 0.1528 \pm 0.0223 $ \\
4402.56562465 	 & 7.00 	 & 6.38 	 & $ 0.1625 \pm 0.0219 $ \\
4402.57542661 	 & 0.10 	 & 6.71 	 & $ 0.1592 \pm 0.0218 $ \\
4402.66299760 	 & 5.99 	 & 7.14 	 & $ 0.1614 \pm 0.0229 $ \\
4402.67279979 	 & -7.23 	 & 7.22 	 & $ 0.1608 \pm 0.0213 $ \\
4403.55353945 	 & -2.78 	 & 5.99 	 & $ 0.1535 \pm 0.0191 $ \\
4403.56276076 	 & -7.91 	 & 6.89 	 & $ 0.1532 \pm 0.0198 $ \\
4403.66105457 	 & 2.66 	 & 7.38 	 & $ 0.1562 \pm 0.0199 $ \\
4403.67027553 	 & -5.76 	 & 7.08 	 & $ 0.1525 \pm 0.0192 $ \\
4405.55611954 	 & -3.48 	 & 7.97 	 & $ 0.1618 \pm 0.0214 $ \\
4405.56534582 	 & -10.03 	 & 7.13 	 & $ 0.1655 \pm 0.0231 $ \\
4570.97352687 	 & 19.43 	 & 8.12 	 & $ 0.1701 \pm 0.0243 $ \\
4604.92111685 	 & 8.57 	 & 8.82 	 & $ 0.1820 \pm 0.0264 $ \\
4605.91057844 	 & 12.66 	 & 7.48 	 & $ 0.1732 \pm 0.0255 $ \\
4662.95425192 	 & 11.46 	 & 8.31 	 & $ 0.1771 \pm 0.0251 $ \\
4662.96610624 	 & -1.38 	 & 8.21 	 & $ 0.1773 \pm 0.0251 $ \\
4663.81542937 	 & 19.18 	 & 9.20 	 & $ 0.1851 \pm 0.0256 $ \\
4663.85662433 	 & 2.50 	 & 7.23 	 & $ 0.1725 \pm 0.0244 $ \\
4663.88867894 	 & 4.10 	 & 6.74 	 & $ 0.1691 \pm 0.0235 $ \\
4663.90391404 	 & 10.16 	 & 6.00 	 & $ 0.1717 \pm 0.0252 $ \\
4665.78413890 	 & 11.48 	 & 7.78 	 & $ 0.1816 \pm 0.0257 $ \\
4665.79430438 	 & -3.10 	 & 9.86 	 & $ 0.1812 \pm 0.0252 $ \\
4665.80583621 	 & 0.63 	 & 9.02 	 & $ 0.1786 \pm 0.0242 $ \\
4731.66600573 	 & 1.60 	 & 11.50 	 & $ 0.1734 \pm 0.0230 $ \\
4731.67754085 	 & 0.72 	 & 7.35 	 & $ 0.1657 \pm 0.0226 $ \\
4731.68907100 	 & -5.82 	 & 8.82 	 & $ 0.1795 \pm 0.0236 $ \\
4732.75396109 	 & 2.27 	 & 7.88 	 & $ 0.1682 \pm 0.0242 $ \\
4732.76548879 	 & 0.12 	 & 7.60 	 & $ 0.1667 \pm 0.0223 $ \\
4732.77701662 	 & -2.03 	 & 6.90 	 & $ 0.1674 \pm 0.0238 $ \\
4733.69703238 	 & 9.44 	 & 8.14 	 & $ 0.1701 \pm 0.0242 $ \\
4733.70856019 	 & -4.67 	 & 7.82 	 & $ 0.1677 \pm 0.0235 $ \\
4733.71697119 	 & -7.09 	 & 9.14 	 & $ 0.1727 \pm 0.0236 $ \\
4747.77353788 	 & 4.25 	 & 5.98 	 & $ 0.1746 \pm 0.0250 $ \\
4781.65688110 	 & 7.76 	 & 6.41 	 & $ 0.1652 \pm 0.0206 $ \\
4816.61232656 	 & 10.29 	 & 8.72 	 & $ 0.1755 \pm 0.0209 $ \\
4986.92370887 	 & -4.73 	 & 14.24 	 & $ 0.1887 \pm 0.0262 $ \\
5025.89140933 	 & -37.17 	 & 8.23 	 & $ 0.1816 \pm 0.0263 $ \\
5048.87005530 	 & -37.84 	 & 9.19 	 & $ 0.1823 \pm 0.0239 $ \\
5048.87535157 	 & -12.18 	 & 9.07 	 & $ 0.1807 \pm 0.0241 $ \\
5075.87456767 	 & -15.99 	 & 7.03 	 & $ 0.1695 \pm 0.0230 $ \\
5075.88610277 	 & -13.42 	 & 8.76 	 & $ 0.1692 \pm 0.0228 $ \\
5100.78063288 	 & -21.91 	 & 7.32 	 & $ 0.1673 \pm 0.0212 $ \\
5135.72812260 	 & -27.54 	 & 8.80 	 & $ 0.1659 \pm 0.0226 $ \\
5172.58815669 	 & -4.45 	 & 8.31 	 & $ 0.1728 \pm 0.0219 $ \\
5400.78427274 	 & -10.86 	 & 6.31 	 & $ 0.1706 \pm 0.0243 $ \\
5435.78861255 	 & -15.49 	 & 8.10 	 & $ 0.1714 \pm 0.0246 $ \\
5469.78414305 	 & -13.65 	 & 7.30 	 & $ 0.1785 \pm 0.0224 $ \\
5492.76854834 	 & -11.30 	 & 7.62 	 & $ 0.1627 \pm 0.0222 $ \\
5523.68425857 	 & 2.79 	 & 7.30 	 & $ 0.1601 \pm 0.0208 $ \\
5527.64465815 	 & 6.66 	 & 8.70 	 & $ 0.1742 \pm 0.0234 $ \\
5527.65392663 	 & 3.10 	 & 5.77 	 & $ 0.1669 \pm 0.0223 $ \\
5547.59458469 	 & 9.04 	 & 8.11 	 & $ 0.1649 \pm 0.0249 $ \\
5644.98896924 	 & 4.16 	 & 6.79 	 & $ 0.1519 \pm 0.0239 $ \\
5722.92872656 	 & 7.82 	 & 5.48 	 & $ 0.1523 \pm 0.0216 $ \\
5758.92483333 	 & 6.08 	 & 6.53 	 & $ 0.1721 \pm 0.0260 $ \\
5789.68097414 	 & -0.49 	 & 6.97 	 & $ 0.1705 \pm 0.0285 $ \\
5791.86463193 	 & 13.74 	 & 8.20 	 & $ 0.1768 \pm 0.0263 $ \\
5792.79925378 	 & -0.60 	 & 6.61 	 & $ 0.1742 \pm 0.0253 $ \\
5841.78442437 	 & -8.29 	 & 7.41 	 & $ 0.1665 \pm 0.0251 $ \\
5842.79215631 	 & -10.22 	 & 6.66 	 & $ 0.1649 \pm 0.0281 $ \\

\end{supertabular}

\end{center}

\begin{center}

\footnotesize

\tablecaption{Radial Velocities for HD 220773}
\label{220773tab}	
\tablefirsthead{\hline
BJD - 2450000 & Radial Velocity & Uncertainty \\
& (m/s) & (m/s) \\
\hline}
		
\tablehead{\hline
\emph{Table \ref{220773tab} cont'd.} & & \\ \hline
BJD - 2450000 & Radial Velocity & Uncertainty \\
& (m/s) & (m/s) \\
\hline}
		
\tabletail{\hline}

\begin{supertabular}{| l l l |}

2479.859725 	 & 	 -0.4171 	         & 	 4.08 	 \\
2480.838237 	 & 	 -13.5865 	 & 	 4.09 	 \\
2486.820898 	 & 	 -11.3423 	 & 	 4.45 	 \\
2487.830499 	 & 	 -16.1442 	 & 	 4.38 	 \\
2948.718426 	 & 	 -10.1010 	 & 	 3.87 	 \\
3185.902803 	 & 	 -24.8888 	 & 	 5.50 	 \\
3267.686065 	 & 	 -22.1946 	 & 	 4.73 	 \\
3542.934309 	 & 	 -35.1965 	 & 	 4.76 	 \\
3604.768270 	 & 	 -26.1322 	 & 	 4.20 	 \\
3901.952807 	 & 	    3.7467 	 & 	 3.70 	 \\
3935.863371 	 & 	  -5.9660 	 & 	 4.19 	 \\
3979.907141 	 & 	 -12.5221 	 & 	 4.13 	 \\
4007.670760 	 & 	    1.0665 	 & 	 3.68 	 \\
4014.657584 	 & 	 -0.4731 	 & 	 3.14 	 \\
4015.639336 	 & 	 5.8017 	 & 	 4.02 	 \\
4032.599555 	 & 	 14.5027 	 & 	 4.60 	 \\
4044.722222 	 & 	 14.3386 	 & 	 4.10 	 \\
4401.744726 	 & 	 12.0180 	 & 	 4.51 	 \\
4428.675456 	 & 	 19.4405 	 & 	 3.78 	 \\
4475.548221 	 & 	 14.3677 	 & 	 4.52 	 \\
5114.635072 	 & 	 -11.1314 	 & 	 4.15 	 \\
5140.721408 	 & 	 7.8337 	 & 	 3.75 	 \\
5358.945550 	 & 	 4.7166 	 & 	 4.79 	 \\
5358.947973 	 & 	 8.1944 	 & 	 4.58 	 \\
5382.886175 	 & 	 -5.2398 	 & 	 3.81 	 \\
5382.888643 	 & 	 -3.2236 	 & 	 4.03 	 \\
5414.822196 	 & 	 -1.2009 	 & 	 3.47 	 \\
5414.824720 	 & 	 -10.5412 	 & 	 3.58 	 \\
5441.730451 	 & 	 -0.5391 	 & 	 3.39 	 \\
5441.732880 	 & 	 -4.1901 	 & 	 3.75 	 \\
5466.827877 	 & 	 -6.5113 	 & 	 4.33 	 \\
5469.822406 	 & 	 -1.8755 	 & 	 3.72 	 \\
5469.824889 	 & 	 -1.5983 	 & 	 3.69 	 \\
5532.649379 	 & 	 -3.1531 	 & 	 4.21 	 \\
5532.651798 	 & 	 8.9105 	 & 	 4.23 	 \\
5547.609328 	 & 	 8.9497 	 & 	 3.68 	 \\
5547.611809 	 & 	 2.8300 	 & 	 4.20 	 \\
5724.945915 	 & 	 9.2505 	 & 	 3.56 	 \\
5724.948402 	 & 	 -5.5916 	 & 	 3.73 	 \\
5754.866046 	 & 	 -12.5525 	 & 	 4.50 	 \\
5754.868528 	 & 	 0.3110 	 & 	 5.85 	 \\
5790.949560 	 & 	 -2.0862 	 & 	 3.38 	 \\
5790.952042 	 & 	 -3.9868 	 & 	 3.78 	 \\

\end{supertabular}

\end{center}

\begin{sidewaystable}

\begin{center}

\clearpage

\footnotesize

\begin{tabular}{l l l l l l l l l l l l l l l}

\hline\hline \\
Star & Spectral & $V$\footnotemark[3] & $B-V$\footnotemark[3] & $M_{V}$ & Parallax\footnotemark[3] & Dist. & T$_{eff}$ & $\log g$ & [Fe/H] & $\xi$ & Mass\footnotemark[4] & Age\footnotemark[4] & $S_{HK}$ & $\log R'_{HK}$ \\ 
       &       Type&        &            &              &(mas)     &(pc)   &(K)         &                &            & (km/s)&(M$_{\odot}$) & (Gyr) \\
\hline \\

HD 79498 & G5 & 8.027 & 0.703 & 4.58 & 20.43 & 49 & $5740 \pm 100$ & $4.37 \pm 0.12$ & $0.24 \pm 0.06$ & $1.27 \pm 0.15$ & 1.06 & 2.70 & 0.167 & -4.66 \\
HD 155358 & G0 & 7.270 & 0.550 & 4.09 & 23.07 & 43 & $5900 \pm 100$ & $4.16 \pm 0.12$ & $-0.51 \pm 0.06$ & $0.50 \pm 0.15$ & 0.92 & 10.70 & 0.169 & -4.54 \\
HD 197037 & F7 V & 6.809 & 0.519 & 4.23 & 30.50 & 33 & $6150 \pm 100$ & $4.40 \pm 0.12$ & $-0.20 \pm 0.06$ & $1.16 \pm 0.15$ & 1.11 & 1.90 & 0.167 & -4.53 \\
HD 220773 & F9 & 7.093 & 0.632 & 3.66 & 20.60 & 49 & $5940 \pm 100$ & $4.24 \pm 0.12$ & $0.09 \pm 0.06$ & $1.50 \pm 0.15$ & 1.16 & 4.40 & 0.159 & -4.98 \\
\hline

\end{tabular}
\caption{Stellar properties.}
\label{stellar}
\end{center}

\footnotetext[3]{\citet{kharchenko09}}
\footnotetext[4]{\citet{casagrande11}}

\end{sidewaystable}

\begin{sidewaystable}

\begin{center}

\footnotesize

\begin{tabular}{l | l | l l | l | l}

\hline\hline & & & & & \\
 & HD 79498b & \multicolumn{2}{| c |}{HD 155358} & HD 197037b\footnotemark[5] & HD 220773b \\
 & & \multicolumn{1}{| c }{b} & \multicolumn{1}{ c |}{c} & & \\
\hline & & & & & \\

Period $P$ (days) & $1966.1 \pm 41$ & $194.3 \pm 0.3$ & $391.9 \pm 1$ & $1035.7 \pm 13$ & $3724.7 \pm 463$ \\
Periastron Passage $T_{0}$ & $3210.9 \pm 39$ & $1224.8 \pm 6$ & $5345.4 \pm 28$ & $1353.1 \pm 86$ & $3866.4 \pm 95$ \\
(BJD - 2 450 000) & & & & \\
RV Amplitude $K$ (m/s) & $26.0 \pm 1$ & $32.0 \pm 2$ & $24.9 \pm 1$ & $15.5 \pm 1$ & $20.0 \pm 3$ \\
Mean Anomaly $M_0$\footnotemark[6] & $329^{\circ} \pm 11^{\circ}$ & $129^{\circ} \pm 0.7^{\circ}$ & $233^{\circ} \pm 0.9^{\circ}$ & $186^{\circ} \pm 3^{\circ}$ & $226^{\circ} \pm 13^{\circ}$ \\
Eccentricity $e$ & $0.59 \pm 0.02$ & $0.17 \pm 0.03$ & $0.16 \pm 0.1$ & $0.22 \pm 0.07$ & $0.51 \pm 0.1$ \\
Longitude of Periastron $\omega$ & $221^{\circ} \pm 6^{\circ}$ & $143^{\circ} \pm 11^{\circ}$ & $180^{\circ} \pm 26^{\circ}$ & $298^{\circ} \pm 26^{\circ}$ & $259^{\circ} \pm 15^{\circ}$ \\
Semimajor Axis $a$ (AU) & $3.13 \pm 0.08$ & $0.64 \pm 0.01$ & $1.02 \pm 0.02$ & $2.07 \pm 0.05$ & $4.94 \pm 0.2$ \\
Minimum Mass $M \sin i$ ($M_{J}$) & $1.34 \pm 0.07$ & $0.85 \pm 0.05$ & $0.82 \pm 0.07$ & $0.79 \pm 0.05$ & $1.45 \pm 0.3$ \\
RMS (m/s) & 5.13 & \multicolumn{2}{c |}{6.14} & 8.00 & 6.57 \\
Stellar ``jitter'' (m/s) & 2.76 & \multicolumn{2}{c |}{2.49} & 2.02 & 5.10 \\
FAP (from periodogram) & $4.09 \times 10^{-9}$ & $1.15 \times 10^{-11}$ & $< 10^{-14}$ & $8.53 \times 10^{-8}$ & $\cdots$  \\
FAP (from bootstrap) & $< 10^{-4}$ & $< 10^{-4}$ & $<10^{-4}$ & $<10^{-4}$ & $\cdots$ \\
\hline
\end{tabular}
\caption{Derived planet properties.}
\label{orbits}
\end{center}

\footnotetext[5]{Parameters include subtraction of linear RV trend in residuals}
\footnotetext[6]{Evaluated at the time of the first RV point reported for each system}

\end{sidewaystable}


\clearpage

\begin{thebibliography}{}
%
\bibitem[Baglin et al.(2003)]{baglin03} Baglin, A. et al. \ 2003, Advances in Space Research, 31, 345
%
\bibitem[G. Benford et al.(2010)]{benfordg10} Benford, G. et al. \ 2010, Astrobiology, 10, 491
%
\bibitem[J. Benford et al.(2010)]{benfordj10} Benford, J. et al. \ 2010, Astrobiology, 10, 475
%
\bibitem[Benz et al.(2008)]{benz08} Benz, W. et al. \ 2008, Physica Scripta Volume T, 130, 014022
%
\bibitem[Bonnarel et al.(2000)]{bonnarel00} Bonnarel, F. et al. \ 2000, \aaps, 143, 33
%
\bibitem[Borucki et al.(2010)]{borucki10} Borucki, W.~J. et al. \ 2010, Science, 327, 977
%
\bibitem[Borucki et al.(2011)]{borucki11} Borucki, W.~J. et al. \ 2011, \apj, 736, 19
%
\bibitem[Boss(2002)]{boss02} Boss, A.~P. \ 2002, \apjl, 567, 149
%
\bibitem[Bromley \& Kenyon(2011)]{bromley11} Bromley, B.~C. \& Kenyon, S.~J. \ 2011, \apj, 731, 101
%
\bibitem[Brugamyer et al.(2011)]{brugamyer11} Brugamyer, E. et al. \ 2011, \apj, 738, 97
%
\bibitem[Brown et al.(2008)]{brown08} Brown, K.~I.~T. et al. \ 2008, \apj, 679, 1531
%
%
\bibitem[Casagrande et al.(2011)]{casagrande11} Casagrande, L. et al. \ 2011, \aap, 530, 138
%
\bibitem[Chambers(1999)]{chambers99} Chambers, J.~E. \ 1999, \mnras, 304, 793
%
\bibitem[Cochran \& Hatzes(1993)]{cochran93} Cochran, W.~D. \& Hatzes, A.~P. \ 1993, ASP Conference Series, 36, 267
%
\bibitem[Cochran et al.(2004)]{cochran04} Cochran, W.~D. et al. \ 2004, \apjl, 611, 133
%
\bibitem[Cochran et al.(2007)]{cochran07} Cochran, W.~D. et al. \ 2007, \apj, 665, 1407
%
\bibitem[Cumming et al.(2008)]{cumming08} Cumming, A. et al. \ 2008, \pasp, 120, 531
%
\bibitem[Del Popoulo et al.(2005)]{delp05} Del Popoulo, A. et al. \ 2005, \aap, 436, 363
%
\bibitem[Demarque et al.(2004)]{demarque04} Demarque, P. et al. \ 2004, \apjs, 155, 667
%
\bibitem[Dodson-Robinson et al.(2009)]{sdr09} Dodson-Robinson, S.~E. \ 2009, Icarus, 200, 672
%
\bibitem[Dommanget \& Nys(2002)]{dommanget02} Dommanget, J. \& Nys, O. \ 2002, VizieR On-line Data Catalog
%
\bibitem[Ecuvillon et al.(2007)]{ecuvillon07} Ecuvillon, A. et al. \ 2007, \aap, 461, 171
%
%
\bibitem[Endl et al.(2000)]{endl00} Endl, M. et al. \ 2000, \aap, 362, 585
%
\bibitem[Fischer \& Valenti(2005)]{fischer05} Fischer, D.~A. \& Valenti, J. \ 2005, \apj, 622, 1102
%
\bibitem[Ford \& Holman(2007)]{ford07} Ford, E.~B., \& Holman, M.~J.\ 2007, \apjl, 664, L51 
%
\bibitem[Fuhrmann \& Bernkopf(2008)]{fuhrmann08} Fuhrmann, K. \& Bernkopf, J. \ 2008, \mnras, 384, 1563
%
\bibitem[Gould et al.(2010)]{gould10} Gould, A. et al. \ 2010, \apj, 720, 1073
%
\bibitem[Horner \& Evans(2006)]{horner06} Horner, J., \& Wyn Evans, N.\ 2006, \mnras, 367, L20
%
\bibitem[Horner \& Jones(2010)]{hornerj10} Horner, J., \& Jones, B.~W.\ 2010, International Journal of Astrobiology, 9, 273
%
\bibitem[Horner \& Lykawka(2010)]{hornerl10} Horner, J., \& Lykawka, P.~S.\ 2010, \mnras, 405, 49 
%
\bibitem[Horner et al.(2011)]{horner11} Horner, J. et al. \ 2011, \mnras, 416, L11
%
%
\bibitem[Jefferys et al.(1988)]{jefferys88} Jefferys, W.~H. et al. \ 1988, Celestial Mechanics, 41, 39
%
%
\bibitem[Johnson et al.(2011)]{johnson11} Johnson, J.~A. et al. \ 2011, \apjs, 197, 26
%
\bibitem[Kasting et al.(1993)]{kasting93} Kasting, J.~F. et al. \ 1993, Icarus, 101, 108
%
\bibitem[Katz et al.(2011)]{katz11} Katz, B. et al. \ 2011, arXiv: 1106.3340
%
\bibitem[Kharchenko \& Roeser(2009)]{kharchenko09} Kharchenko, N.~V. \& Roeser, S. \ 2009, VizieR On-line Data Catalog
%
%
\bibitem[Kozai(1962)]{kozai62} Kozai, Y. \ 1962, \aj, 67, 591
%
\bibitem[K\"{u}rster et al.(1997)]{kurster97} K\"{u}rster, M. et al. \ 1997, \aap, 320, 831
%
\bibitem[Kurucz(1993)]{kurucz93} Kurucz, R. \ 1993, ATLAS9 Stellar Atmosphere Programs and 2 km/s grid.  Kurucz CD-ROM No. 13 (Cambridge: Smithsonian Astrophys. Obs.)
%
\bibitem[Lambert et al.(1991)]{lambert91} Lambert, D.~L. et al. \ 1991, \mnras, 349, 757
%
\bibitem[Lammer et al.(2009)]{lammer09} Lammer, H. et al. \ 2009, Astron. Astrophys. Rev., 17, 181
%
\bibitem[Latham et al.(1989)]{latham89} Latham, D.~W. et al. \ 1989, \nat, 339, 8
%
\bibitem[Lidov(1962)]{lidov62} Lidov, M.~L. \ 1962, \ Planetary and Space Science, 9, 719
%
\bibitem[Lykawka et al.(2009)]{lykawka09} Lykawka, P.~S., Horner, J., Jones, B.~W., \& Mukai, T.\ 2009, \mnras, 398, 1715 
%
\bibitem[Lykawka \& Horner(2010)]{lykawka10} Lykawka, P.~S., \& Horner, J.\ 2010, \mnras, 405, 1375 
%
\bibitem[Lykawka et al.(2011)]{lykawka11} Lykawka, P.~S., Horner, J., Jones, B.~W., \& Mukai, T.\ 2011, \mnras, 412, 537 
%
%
\bibitem[Mandell et al.(2007)]{mandell07} Mandell, A.~M. et al. \ 2007, \apj, 660, 823
%
\bibitem[Mann et al.(2010)]{mann10} Mann, A.~W. et al. \ 2010, \apj, 719, 1454
%
\bibitem[Marshall et al.(2010)]{marshall10} Marshall, J. et al. \ 2010, International Journal of Astrobiology, 9, 259
%
\bibitem[Mayor \& Queloz(1995)]{mayor95} Mayor, M. \& Queloz, D. \ 1995, \nat, 378, 355
%
\bibitem[Meschiari et al.(2009)]{meschiari09} Meschiari, S. et al. \ 2009, \pasp, 121, 1016
%
\bibitem[Morbidelli et al.(2005)]{morbidelli05} Morbidelli, A., Levison, H.~F., Tsiganis, K., \& Gomes, R.\ 2005, \nat, 435, 462 
%
\bibitem[Nelan et al.(2010)]{nelan10} Nelan, E. et al. \ 2010, ``Fine Guidance Sensor Instrument Handbook,'' Version 18.0, (Baltimore: STScI)
%
\bibitem[Noyes et al.(1984)]{noyes84} Noyes, R.~W. et al. \ 1984, \apj, 279, 763
%
\bibitem[Porter \& Grundy(2011)]{porter11} Porter, S.~B. \& Grundy, W.~M. \ 2011, \apjl, 736, 14
%
\bibitem[Queloz et al.(2001)]{queloz01} Queloz, D. et al. \ 2001, \aap, 379, 279
%
\bibitem[Ramsey et al.(1998)]{ramsey98} Ramsey, L.~W. et al. \ 1998, Proc. SPIE, 3352, 34
%
%
\bibitem[Schulze-Makuch et al.(2011)]{dsm11} Schulze-Makuch, D. et al. \ 2011, Astrobiology, 11, Pre-Print
%
%
\bibitem[Setiawan et al.(2011)]{setiawan11} Setiawan, J. et al. \ 2011, \nat, 330, 1642
%
\bibitem[Sneden(1973)]{sneden73} Sneden, C.~A. \ 1973, Ph.D. thesis, Univ. of Texas at Austin
%
\bibitem[Sturrock \& Scargle(2010)]{sturrock10} Sturrock, P.~A. \& Scargle, J.~D. \ 2010, \apj, 718, 527
%
\bibitem[Sumi et al.(2011)]{sumi11} Sumi, T. et al. \ 2011, \nat, 473, 349
%
\bibitem[Tinney et al.(2011)]{tinney11} Tinney, C.~G. et al. \ 2011, \apj, 732, 31
%
\bibitem[Tsiganis et al.(2005)]{tsiganis05} Tsiganis, K. et al. \ 2005, \nat, 435, 459
%
\bibitem[Tull et al.(1995)]{tull95} Tull, R.~G. et al. \ 1995, \pasp, 107, 251
%
\bibitem[Tull et al.(1998)]{tull98} Tull, R.~G. et al. \ 1998, Proc. SPIE, 3355, 387
%
\bibitem[Udry \& Santos(2007)]{udry07} Udry, S. \& Santos, N.~C. \ 2007, \araa, 45, 397
%
%
\bibitem[Wittenmyer et al.(2007)]{witt07} Wittenmyer, R.~A., Endl, M., Cochran, W.~D., \& Levison, H.~F.\ 2007, \aj, 134, 1276 
%
\bibitem[Wittenmyer et al.(2009)]{witt09} Wittenmyer, R.~A., Endl, M., Cochran, W.~D., Levison, H.~F., \& Henry, G.~W.\ 2009, \apjs, 182, 97 
%
\bibitem[Wittenmyer et al.(2011a)]{witt11} Wittenmyer, R.~A. et al. \ 2011a, \apj, 727, 102
%
\bibitem[Wittenmyer et al.(2011b)]{witt11b} Wittenmyer, R.~A. et al. \ 2011b, arXiv: 1110.2542
%
\bibitem[Wright et al.(2011)]{wright11} Wright, J.~T. et al. \ 2011, \pasp, 123, 412
%
\bibitem[Zechmeister \& K\"{u}rster(2009)]{zechmeister09} Zechmeister, M. \& K\"{u}rster, M. \ 2009, \aap, 496, 577
%
\end{thebibliography}
\end{document}